\documentclass[12pt]{article}
\usepackage{amsmath}
\usepackage{graphicx}
\usepackage{enumerate}
\usepackage{natbib}
\usepackage{url} 
\usepackage{multibib}

\newcites{Supp}{Supplementary References}

\usepackage{algorithm}
\usepackage{algpseudocode}  
\usepackage{esvect}
\usepackage{multirow}
\usepackage{xcolor}
\usepackage{booktabs}
\usepackage[caption=false,font=footnotesize]{subfig}
\usepackage{amssymb}
\usepackage{MnSymbol}

\newtheorem{theorem}{Theorem}[section]

\newtheorem{proposition}{Proposition}[section]

\newtheorem{remark}{Remark}[section]

\newtheorem{lemma}{Lemma}[section]

\newtheorem{definition}{Definition}[section]

\newtheorem{algo}{Task}

\newcommand{\bX}{{\bf X}}

\newcommand{\bea}{\begin{eqnarray}}
\newcommand{\eea}{\end{eqnarray}}
\newcommand{\barr}{\begin{array}}
\newcommand{\earr}{\end{array}}
\newcommand{\bn}{\begin{enumerate}}
\newcommand{\en}{\end{enumerate}}
\newcommand{\ei}{\end{itemize}}
\newcommand{\bbbm}{\begin{pmatrix}}
\newcommand{\eeem}{\end{pmatrix}}

\newcommand{\bbf}{{\bf f}}
\newcommand{\bbg}{{\bf g}}

\newcommand{\bbm}{{\bf m}}

\newcommand{\bbv}{{\bf v}}

\newcommand{\bbx}{{\bf x}}
\newcommand{\bby}{{\bf y}}

\newcommand{\bbG}{{\bf G}}

\newcommand{\bbR}{{\bf R}}

\newcommand{\bbU}{{\bf U}}

\newcommand{\bbX}{{\bf X}}
\newcommand{\bbY}{{\bf Y}}

\newcommand{\cA}{{\cal A}}

\newcommand{\cC}{{\cal C}}
\newcommand{\cD}{{\cal D}}

\newcommand{\cH}{{\cal H}}

\newcommand{\cP}{{\cal P}}

\newcommand{\cL}{{\cal L}}

\newcommand{\E}{{\mathbb E}}

\newcommand{\N}{{\mathbb N}}

\newcommand{\R}{{\mathbb R}}

\newcommand{\ga}{\gamma}
\newcommand{\Ga}{\Gamma}
\newcommand{\de}{\delta}
\newcommand{\De}{\Delta}

\newcommand{\la}{\lambda}

\newcommand{\vp}{\varphi}

\newcommand{\si}{\sigma}

\newcommand{\ignore}[1]{}{}

\newcommand{\nn}{\nonumber}

\newcommand{\p}{{\partial}}

\newcommand{\wh}{\widehat}

\newcommand{\q}{\quad}

\newcommand{{\QED}}{{\hfill QED} \bigskip}

\DeclareMathSymbol{\shortminus}{\mathbin}{AMSa}{"39}

\newcommand{\eps}{\varepsilon}

\newcommand{\xy}{\langle x,  y \rangle}

\newcommand{\prox}{\ensuremath{\operatorname{Prox}}}
\newcommand*{\trans}{\mathsf{T}}

\DeclareMathOperator{\id}{Id}

\DeclareMathOperator*{\argmin}{argmin}

\DeclareMathOperator*{\Var}{{Var}}
\DeclareMathOperator*{\Cov}{{Cov}}

\definecolor{darkspringgreen}{rgb}{0.09, 0.45, 0.27} 
\definecolor{darkgray}{rgb}{0.66, 0.66, 0.66}

\begin{document}

\def\spacingset#1{\renewcommand{\baselinestretch}%
{#1}\small\normalsize} \spacingset{1}


  \title{\bf Monotone Curve Estimation via Convex Duality}
  \author{Tongseok Lim\thanks{
The authors gratefully acknowledge the support of the National Research Foundation of Korea (RS-2019-NR040050).
}\hspace{.2cm}\\
    Mitch Daniels School of Business, Purdue University, USA\\ 
    Kyeongsik Nam\\
    Department of Mathematical Sciences, KAIST, South Korea\\
    and \\
    Jinwon Sohn \\
    Department of Statistics, Purdue University, USA}
  \maketitle


\bigskip
\begin{abstract}
A principal curve serves as a powerful tool for uncovering underlying structures of data through 1-dimensional smooth and continuous representations. On the basis of optimal transport theories, this paper introduces a novel principal curve framework constrained by monotonicity with rigorous theoretical justifications. We establish statistical guarantees for our monotone curve estimate, including expected empirical and generalized mean squared errors, while proving the existence of such estimates. These statistical foundations justify adopting the popular early stopping procedure in machine learning to implement our numeric algorithm with neural networks. Comprehensive simulation studies reveal that the proposed monotone curve estimate outperforms competing methods in terms of accuracy when the data exhibits a monotonic structure. Moreover, through two real-world applications on future prices of copper, gold, and silver, and avocado prices and sales volume, we underline the robustness of our curve estimate against variable transformation, further confirming its effective applicability for noisy and complex data sets. We believe that this monotone curve-fitting framework offers significant potential for numerous applications where monotonic relationships are intrinsic or need to be imposed.
\end{abstract}

\noindent%
\vfill

\newpage
\spacingset{1.9} 
\section{Introduction}
\label{sec:intro}

A principal curve is a 1-dimensional manifold penetrating the middle of data. In mathematics, a general curve is defined as a vector of functions where an index parameter draws the shape of each component function. For example, we can parametrize a unit circle as $(\cos s, \sin s)$, or a quadratic curve in $\mathbb{R}^2$ as a vector of $(s,s^2)$, where $s \in \R$ denotes the index parameter. Similarly, a principal curve is a 1-dimensional manifold of multivariate random variables whose probability law relies on the index parameter. A representative example is a multivariate Gaussian distribution whose mean vector varies by the index parameter. In this regard, a principal curve captures the essential geometric shape of data, and it also provides curve-sense summary statistics to analyze joint relationships between variables rather than conditional relationships. The latter characteristic of a principal curve can be particularly beneficial when causal relationships between variables are ambiguous, which occurs a lot in numerous real-world problems. 

Various fields harness such characteristics of a principal curve in data analysis. \cite{banf:raft:92} clustered principal curves to outline distinctive floes in a satellite image as an image processing method. In the field of transportation, \cite{einb:dwye:11} analyzed the relationship between vehicles' speed and traffic flow by curves since the two key variables are hardly seen to have a causal relationship with each other but rather a latent variable traffic density associates the target variables. Also, for pathway-level analysis on cancer progression, \cite{drie:etal:13} projected normal and tumor samples on a principal curve estimated from the principal components of the two contrasting samples and showed the distance along the curve between the samples reveals important findings on cancer data analysis.  \cite{chen:etal:15} employed a principal curve technique to detect and compare cosmic filaments from densities of the galaxy and dark matter. For more examples of using principal curves for data analysis, we refer to   \cite{take:etal:21,layt:etal:22,ghaz:etal:24}.

While the literature on principal curves becomes more mature, the concept of a principal curve constrained to exhibit a monotonic shape remains underexplored. Monotonic relationships between variables are prevalent in numerous real-world scenarios, addressing the practical importance of estimating a monotone principal curve for reliable analysis. For instance, demand and supply curves in economics illustrate inverse and direct relationships between price and quantity. \cite{arfa:etal:17} analyzed the interdependencies between oil, gold, US dollars, and the stock market through a simultaneous equation model, which justifies that each variable relates positively or negatively to other variables. As addressed in \cite{patt:etal:10}, a plethora of theories in financial modeling require monotonic situations such as discovering the relation between expected returns and market betas in the capital asset pricing model. In ethics-considered machine learning, such as for privacy or fairness,  ethical improvement of a learned model tends to trade off its utility such as degree of accuracy in general \citep{xie:etal:18,sohn:etal:24}. These real-world problems highlight the potential utility of monotone principal curves as fundamental data analysis tools when monotonic phenomena between variables are expected.

Our ontributions are mainly threefold. First, we propose a novel principal curve framework for fitting a monotonic curve in Section~\ref{sec:loss}. To begin with, we characterize the monotone curve in arbitrary finite-dimensional space using convex analysis and monotone operator theory, and then we propose an optimization problem to find the monotone curve. Secondly, Section~\ref{sec:error_analysis} places rigorous theoretical analysis on the curve estimate, the solution of the established optimization. After verifying the existence of the solution, we disclose the convergence rate of the expected empirical and generalization errors. The generalization error in this work implies the estimation error for unseen data. Note that tackling such a generalization error is particularly important from machine learning and statistics viewpoints, since it can provide a theoretical foundation for model selection to avoid overfitting or find optimal hyperparameters. To our knowledge, this is the first work in the principal curve literature to derive the generalization error based on convex analysis. Finally, in Section~\ref{sec:numeric}, we introduce a numerical algorithm that finds the monotone curve through neural networks. In Section~\ref{sec:simulation}, simulation studies verify that our method achieves more accurate estimation than general curve-fitting methods when the underlying curve is monotone. Section~\ref{sec:realdata} applies the proposed framework to model future prices for copper, gold, and silver, and to estimate a demand curve between avocado prices and sales volume. These real data applications advertise that our method enjoys more robust estimation in terms of variable transformation than the competing methods when monotone relationships are anticipated.

\subsection{Literature review}
\label{sec:literature_curve}

To seek a better representation of a principal curve, versatile approaches have been steadily attempted and categorized mainly by two branches from our perspective.  First, a principal curve is seen as a minimizer of a distance between the curve and data points with regularization. Secondly, a principal curve is approached by a ridge where gradients and eigenvectors of a density's Hessian are orthogonal. The seminal work of \cite{hast:stue:89} defined a curve on self-consistency as a local average of data points having the minimal mean squared distance to the data points indexed by the domain of the curve itself. This initial approach, however, guarantees the existence of a curve only on a specific data distribution and tends to estimate a biased curve around high-curvature areas. These limitations ignited subsequent studies. To alleviate the bias, \cite{tibs:92} approached a principal curve as weighted local averages of data points by formulating the curve-generating process as a probabilistic mixture modeling. \cite{banf:raft:92} eased the bias by updating a curve estimate based on the smoothed residuals of the curve. To ensure the existence of a curve in a more general distribution, \cite{kegl:etal:00} considered a curve whose length is bounded  by a fixed size on a bounded and convex domain. \cite{gerb:whit:13} proposed a surrogate objective to replace the mean squared distance so that a principal curve lies in critical points that are locally minimal. In behind, \cite{duch:stue:96} showed the self-consistent curve can be critical but not extremal points for the mean squared distance, which contributed to justifying such challenges of estimating self-consistent curves. Recent efforts in this branch have focused on finding self-consistent curves in non-Euclidean spaces \citep{haub:15,lee:etal:20,kang:oh:24}.

On the one hand, \cite{ozer:etal:11} viewed a principal curve, for a given probability density function, to lie in an 1-dimensional intrinsic space in which the density's gradient and eigenvectors of the Hessian are constrained to be orthogonal. The set of points in the constrained space is called a ridge or filament that captures the essential structures of the data \citep{hall:92,eber:94,qiao:polo:16}. In general, ridges refer to multidimensional constrained spaces where the projected gradient of the density vanishes in certain directions defined by the Hessian eigenvectors, and the curvature satisfies specific eigenvalue constraints. Note the filament stands for the 1-dimensional ridge. \cite{geno:etal:14} mathematically justified that this subspace-constrained approach that finds a ridge can approximate the true lower-dimensional manifold with an additive noise term. \cite{chen:etal:15:2} suggested selecting a reasonable kernel's bandwidth used to approximate the probability density by maximizing the extent of covering a population ridge. On the other hand, \cite{qiao:polo:16} viewed the 1-dimensional ridge as an integral curve from the differential equation involving the Hessian's eigenvector corresponding to the second eigenvalue in the two-dimensional space. Besides, further efforts have been made to improve the algorithmic computation and theory of the subspace-constrained methodology \citep{qiao:polo:21, zhan:chen:23}.

\section{Formulation of monotone curve-fitting task}
\label{sec:loss}

The construction of our monotone curve-fitting task starts with characterizing a monotone curve in a finite-dimensional space through convex analysis. Then a novel learning problem is introduced where the solution corresponds to the suggested monotone curve.

\subsection{Monotone set and diagonal parametrization}
\label{sec:monoset}

Let $[n]:= \{1,2,...,n\}$ for $n \in \N = \{1,2,...\}$. Let $\cP(\Omega)$ denote the set of probability measures (distributions) over a measure space $\Omega$, and $\langle {\bf a}, {\bf b} \rangle$ denote the inner product of vectors ${\bf a},{\bf b}$. We say that a set $\Ga \subset \R^k$ is  \emph{monotone} if for any ${\bf a} = (a_i)_i, {\bf b} = (b_i)_i \in \R^k$, either $a_i \ge b_i$ or $a_i \le b_i$ for all $i \in [k]$. Note that this is equivalent to $(b_j - a_j)(b_i - a_i) \ge 0$ for all $i, j \in [k]$. A monotone set $\Ga$ is {\em maximally monotone} if it is not a proper subset of other monotone sets.

Now we characterize a monotone curve via the diagonal coordinate $s(\bbx) := \sum_{i=1}^k x_i, \forall \bbx = (x_1,...,x_k) \in \R^k$. Let $S(\Ga) := \{ s(\bbx) \, | \, \bbx \in \Ga\}$ and $s_\Ga : \Ga \to S(\Ga)$, the restriction of $s$ on $\Ga$. 
\begin{definition}
For a monotone set $\Ga \subseteq \R^k$, we call $\ga= (\ga_1,...,\ga_k) :=s_\Ga^{-1}: S(\Ga) \to \Ga$ the associated {\em monotone curve}, which parametrizes $\Ga$ via the diagonal coordinate $s$. 
\end{definition}
Note that $s_\Ga$ is bijective and each component $\ga_i$ is a nondecreasing function of $s$ if $\Ga$ is monotone, and the maximal monotonicity of $\Ga$ is equivalent to the condition $S(\Ga) = \R$. Our definition of a monotone curve is based on an underlying monotone set. In the following, we outline the characterization of a monotone set using convex functions that satisfy a duality relationship, and show that the associated curve $\gamma$ can be represented by them.

\subsection{Exposure of a monotone set through convex functions} 

We first introduce basic mathematical tools for convex analysis. Let ${\cal A}(\cH)$ denote the set of proper, lower-semicontinuous and convex functions (valued in $\R \cup \{+\infty\}$) on a Hilbert space $\cH$.  In this general setting, a set $\Gamma \subset \cH \times \cH$ is called monotone if $\langle b_2 - a_2, b_1 - a_1 \rangle \ge 0$  for any ${\bf a} = (a_1,a_2), {\bf b} = (b_1,b_2)  \in \Gamma$.  For a function $f: \cH \to \R \cup \{+\infty\}$, its convex conjugate is  $f^*(\bby) := \sup_{\bbx \in \cH}  \, [\langle \bbx, \bby \rangle - f(\bbx)]$. Then the Fenchel-Young inequality $f(\bbx) + f^*(\bby) \ge \langle \bbx, \bby \rangle$ holds for all $\bbx,\bby \in \cH$. We say that $f$ and $g$ are  \emph{mutually conjugate} if $f = g^*$ and $g = f^*$. The Fenchel–Moreau theorem states that  $f^{**} := (f^*)^* = f$ for any $f \in \cA(\cH)$, which implies $f$ and $f^*$ are mutually conjugate. For $f \in \cA(\cH)$ and $\bbx \in \cH$, the \emph{subdifferential} of $f$ at $\bbx$ is defined as the following convex set $\p f (\bbx) = \{ \bbv \in \cH \mid f(\bby) -f(\bbx) \ge \langle \bbv, \bby-\bbx \rangle \, \forall \bby \in \cH\}$, and the set $\displaystyle \p f = \bigcup_{\bbx \in \cH} \{(\bbx,\bby) \in \cH \times \cH \mid \bby \in \p f (\bbx) \} $ is then called the subdifferential of $f$.

The following proposition by Rockafellar and Minty (see \cite{bauschke2019convex}) states that a monotone set is contained in a set where the Fenchel-Young inequality achieves equality. Let $ S(\Gamma) = \{ \bbx+\bby \mid (\bbx,\bby) \in \Gamma \}$ and $\Gamma^{-1} := \{(\bby,\bbx) \mid (\bbx, \bby) \in \Gamma \}$ for $\Gamma \subset \cH^2$.
\begin{proposition}\label{classicconvex}
Let $f, g \in \cA(\cH)$ satisfy $f(x) + g(y)  \ge \xy $ for all $x,y \in \cH$. Then the contact set $\Gamma_{f,g} := \{(x,y) \in \cH^2 \ | \ f(x) + g(y) = \xy \}$ is monotone. Moreover, $f, g$ are mutually conjugate if and only if $\Gamma_{f,g}$ is maximally monotone if and only if $S(\Gamma_{f,g}) = \cH$, in which case $
\Gamma_{f,f^*} = \p f =(\p f^*)^{-1}$. Furthermore, any monotone set $\Gamma \subset \cH^2$ is contained in a maximally monotone set $\Gamma_{f,f^*}$ for some $f \in \cA(\cH)$.
\end{proposition}
The equality $\p f =(\p f^*)^{-1}$ states that for $f \in \cA(\cH)$, $\p f$ and $\p f^*$ are the (generalized) inverses of each other. 

Understanding how to obtain such $f$ and $f^*$ that expose a given monotone set $\Ga$ as $\Ga \subset \Ga_{f,f^*}$ is particularly simple and instructive when $\cH = \R$. To simplify discussion, we suppose that $\Ga \subset \R^2$ is monotone and $\Ga$ is maximally monotone without loss of generality.

\begin{remark}[Construction of $f,f^*$ given $\Ga$]\label{construction2}
Given a maximally monotone $\Ga \subset \R^2$, we define convex functions $f,g$ such that $H_{f,g} (x,y):= f(x) + g(y) - xy \ge 0$ for all $x,y \in \R$, and moreover, $\Ga = \{ (x,y) \in \R^2 \, | \, H_{f,g}(x,y) = 0 \}$. As a result, $f,g$ are mutually conjugate due to the maximality of $\Ga$.
For simplicity, we assume $\Ga$ is strictly monotone, i.e., $(x'-x)(y'-y) > 0$ for any $(x,y), (x',y') \in \Ga$. We also assume that $\{x \, | \, (x,y) \in \Ga\} = \{y \, | \, (x,y) \in \Ga \} = \R$. Then $\Ga$ is the graph of a continuous, strictly increasing function (still denoted as $\Ga$), i.e., $\Ga = \{(x,y) \mid y = \Ga(x),\, x \in \R \}$. Also, $\Ga$ has a unique intersection point with $y$-axis; denote it as $(0, y_0)$. Now define $f(x):= \int_0^x \Ga(u)du$ and $g(y) := \int_{y_0}^y \Ga^{-1}(u)du$. Then $f,g$ are convex as $\Ga$ is increasing, and $\p g = (\p f)^{-1}$ implies that $f,g$ are mutually conjugate, i.e., $g=f^*$.
\end{remark}

\subsection{Loss function to extract a monotone set}
\begin{figure}[h!]
    \centering
    \includegraphics[width=1.0\linewidth]{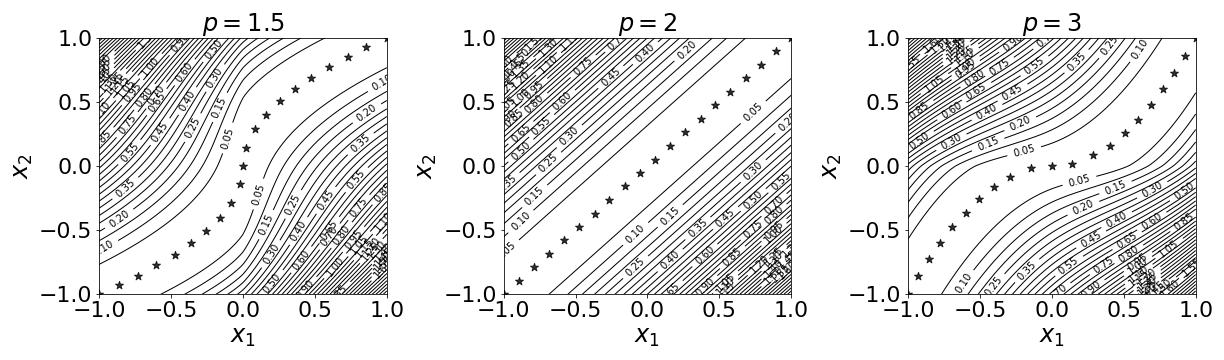}
    \caption{Contour of $H(\bbx;f,f^*)$: $f(x)=|x|^p/p$ and $f^*(x)=|x|^q/q$, the convex conjugate of $f$, with $1/p+1/q=1$. The star points ($\filledstar$) represent the zero set where $H(\bbx;f,f^*)=0$. }
    \label{fig:toy_value_map}
\end{figure}

Such procedures for characterizing a monotone set through mutually conjugate convex functions lay the groundwork for designing a novel statistical learning framework to identify a monotone principal curve. It is helpful to first discuss this problem in the two-dimensional space \(\R^2\). Let \(\mu \in \cP(\R^2)\) denote the data distribution of \(\bbX = (X_1, X_2)\), which exhibits an approximately monotone structure. To formulate a learning framework, this work proposes the following “loss function” to identify a monotone set:
\begin{align*}
H(\bbx; f_1,f_2) = f_1(x_1) + f_2(x_2) - x_1 x_2, \q \bbx = (x_1,x_2) \in \R^2
\end{align*}
where $f_1$ and $f_2$ are in duality defined as below.  
\begin{definition}
    We say that a pair of convex functions $f_1, f_2 \in \cA(\R)$ is in {\em duality position}, or simply {\em in duality}, if $H(\bbx; f_1,f_2) \ge 0$, i.e., $f_1(x_1) + f_2(x_2)  \ge x_1x_2 $ for all $x_1, x_2 \in \R$. 
\end{definition}
This loss function will be generally larger as $\bbx$ is farther from the monotone set $\Gamma_{f_1,f_2}$ on which $H$ attains its minimum $0$. Figure~\ref{fig:toy_value_map} draws a contour of $H(\bbx; f,f^*)$ for a power function on the grid of $x_1,x_2$.
This inspires us to use $\E_{\bbX \sim \mu} [H(\bbX; f_1,f_2)]$ as an objective and minimize it over the convex pair $f_1,f_2$ in duality, since Proposition \ref{classicconvex} shows that the zero set \(\Gamma_{f_1,f_2}\) will then represent a monotone curve in \(\mathbb{R}^2\). Later, we will present a generalized formulation for higher dimensions \(\mathbb{R}^k\) in \eqref{problem3d} and \eqref{problem3d-1}.

We now explain how the diagnoal parametrization (monotone curve) $\ga = s_\Ga^{-1} : S(\Gamma) \to \Gamma$ can be represented by a convex pair $(f_1,f_2)$. If $(f_1, f_2)$ is in duality and $\bbx = (x_1,x_2)$, the first order condition yields the following implication
\begin{align}\label{foc2}
H(\bbx; f_1,f_2) = 0 \implies x_2 \in \p f_1 (x_1) \, \text{ and } \, x_1 \in \p f_2(x_2). 
\end{align}
With $q(x) := |x|^2/2$ and $g_i := f_i + q$,\, $i=1,2$, the following equivalences $x_2 \in \p f_1 (x_1) \iff x_1 + x_2 \in \p g_1 (x_1)$ and $x_1 \in \p f_2(x_2) \iff x_1 + x_2 \in \p g_2 (x_2)$ are straightforward. Hence, the implication \eqref{foc2} can be restated as $\bbx \in \Ga_{f_1,f_2} \implies x_i = (\p g_i)^{-1} (s(\bbx)), \ i=1,2$. In fact, the reverse implication also holds; see \cite[Theorem 4.1]{bartz2021multi} and its proof. 

This shows that the diagonal parametrization $\ga =(\ga_1,\ga_2)$ of the monotone set $\Ga_{f_1,f_2}$ can be expressed as $\ga_i = (\p g_i)^{-1}$. Moreover, $(\p g_i)^{-1} = \nabla g_i^*$, where (whether $g_i \in \cA(\R)$ is differentiable or not) $g_i^* = f_i^* \square q := \inf_{y \in \R} [ f_i^*(y) + q (x-y) ]$, called the Moreau envelope of $f_i^*$, is a differentiable convex function on $\R$ with $1$-Lipschitz nondecreasing derivative $\nabla g_i^*$ (\cite{bauschke2019convex}). Hence the monotone curve $\ga(s) := \big((\p g_1)^{-1}(s), (\p g_2)^{-1}(s)\big)$ is defined for all $s \in \R$, with $\big((\p g_1)^{-1}(s), (\p g_2)^{-1}(s)\big) \in \Ga_{f_1,f_2}$ if $s \in S(\Ga_{f_1,f_2})$. In particular, if $f_1 = f = f_2^*$, then since $S(\Ga_{f, f^*}) = \R$, $\ga$ yields a bijection between $\R$ and $\Ga_{f, f^*}$.  

\subsection{Exposure of a monotone curve in higher dimensions}
 This section extends the aforementioned discussion to any data dimension $k$.
 For $f_i \in \cA(\R)$, $i \in [k]$, we write $\bbf = (f_1,...,f_k)$ and $f^\oplus (\bbx) := \sum_{i=1}^k f_i(x_i)$ for $\bbx = (x_1,...,x_k) \in \R^k$. Define a cost function $c(\bbx)  := \sum_{1\leq i<j\leq k} x_i x_j$ and $H(\bbx; \bbf) := f^\oplus (\bbx) - c(\bbx) =  \sum_{i=1}^k f_i (x_i) - \sum_{1\leq i<j\leq k} x_i x_j$. Let us say that {\em $\bbf$ is in duality position}, or simply {\em in duality}, if the nonnegativity $H(\bbx; \bbf) \ge 0$ holds for all $\bbx \in \R^k$. Also we say that $\bbf = (f_1,\ldots,f_k)$ is a {\em $c$-conjugate tuple} if for each $ i \in [k]$ and $x_{i} \in \R$, $f_{i}(x_{i})=(\bigoplus_{j \neq i}f_j )^c(x_i) :=\sup_{j \neq i,\, x_j \in \R}\  [ c(x_1, \dots,x_i, \dots, x_k)-\sum_{j \neq i}f_j (x_j)]$. Any $c$-conjugate tuple is in duality position. Also, $(\bigoplus_{j \neq i}f_j )^c$ is convex and lower-semicontinuous for any proper (not necessarily convex) functions $(f_j)_j$, thus $(\bigoplus_{j \neq i}f_j )^c \in \cA(\R)$ if proper.
\begin{remark}\label{fg}
Let $q(x) = x^2/2$, $g_i = f_i + q$. We then have $H(\bbx; \bbf) = \widetilde H(\bbx; \bbg)$ with $\widetilde H(\bbx; \bbg) := g^\oplus (\bbx) - \tilde c(\bbx)$ and $\tilde c(\bbx) := s(\bbx)^2/2$. As a result, $\bbf$ is $c$-conjugate if and only if $\bbg$ is $\tilde c$-conjugate.
\end{remark}

The following is an extension of Proposition \ref{classicconvex} for the $c$-conjugate functions on $\R$. 

\begin{proposition}\label{multiconvex}
 Assume that $\bbf = (f_i)_{i=1}^k \in \cA(\R)^k$ is in duality position. Then the zero set 
$\Gamma_{\bbf} := \{\bbx \in \R^k \, | \, H(\bbx; \bbf) = 0 \}$ is monotone. Moreover, $\bbf$ is a $c$-conjugate tuple if and only if $\Gamma_{\bbf}$ is maximally monotone if and only if $S(\Gamma_{\bbf}) = \R$. Furthermore, any monotone set $\Gamma \subset \R^k$ is contained in a maximally monotone set $\Gamma_{\bbf}$ for some $c$-conugate tuple $\bbf \in \cA(\R)^k$.
\end{proposition}

The monotonicity of $\Ga_{\bbf}$ is shown in Lemma \ref{lemma1} in Supplementary~\ref{supp:proof}. The proposition shows that any monotone curve in $\R^k$ can be exposed as a subset of the monotone set $\Gamma_{\bbf}$ induced by convex functions $\bbf$ in duality, and  $\Gamma_{\bbf}$ is maximally monotone if and only if $\bbf$ is $c$-conjugate. This motivates us to use $H(\bbx; \bbf)$ as the objective function. We note that Remark~\ref{classicconvex} can also be extended for $\R^k$. This is outlined in Supplementary~\ref{supp:extra}.

To characterize a monotone curve $\gamma$ in $\R^k$, we turn to the parametrization of monotone sets in $\R^k$ using the diagonal coordinate. Given $f_i \in \cA(\R)$, $i \in [k]$, with $H(\, \cdot \,; \bbf) \ge 0$ and $\bbx = (x_1,...,x_k) \in \R^k$, the  equivalence $H(\bbx; \bbf) = 0 \iff s(\bbx) \in \p g_i (x_i)  \ \forall i \in [k]$ holds true by an application of Theorem 4.1 of \cite{bartz2021multi}.  This yields a parametrization of the monotone set $\Ga_{\bbf}$ via the diagonal coordinate, i.e., $\bbx \in \Ga_{\bbf} \iff x_i = (\p g_i)^{-1} (s(\bbx)), \ \forall i \in [k]$, where $(\p g_i)^{-1} = \nabla g_i^* = \nabla (f_i^* \square q)$ is a $1$-Lipschitz nondecreasing function on $\R$. We can thus write $\ga(s) = \ga_{\bbf}(s) := \big((\p g_i)^{-1}(s)\big)_{i \in [k]}$. If $\bbf$ is $c$-conjugate, then $\ga : \R \to \Ga_{\bbf}$ is a bijection.

\subsection{Monotone curve-fitting task}

 The diagonal parametrization motivates another natural form for the objective function \(\| \bbx - \ga_{\bbf} (s(\bbx)) \|^2\). Intuitively, $\ga_{\bbf}$ minimizing the squared error can be seen as a 1-dimensional latent manifold that condenses the original information of the data points. This extra term, therefore, helps $\ga_{\bbf}$ represent the essential geometric shape of the data points. In manifold learning, this term is commonly referred to as a reconstruction error, representing the extent to which a model can accurately reproduce the original data. Consequently, we propose the following optimization problem for the monotone curve-fitting task (for a chosen parameter $\la \ge 0$ and a domain $\cD \subset \cA(\R)$):
\begin{align}\label{problem3d}
\text{minimize }\, \E_{\bbX \sim \mu} \big[H(\bbX; \bbf) + \la  \| \bbX - \ga_{\bbf} (s(\bbX)) \|^2 \big]  \text{ over }  \bbf = (f_1,...,f_k) \in \cD^k \text{ in duality.}
 \end{align}
As before, if we restrict $\bbf$ to be $c$-conjugated, then the problem is expressed as 
\begin{align}\label{problem3d-1}
\text{minimize }\, \E_{\bbX \sim \mu} \big [H(\bbX; \bbf) + \la \| \bbX - \ga_{\bbf} (s(\bbX)) \|^2  \big] 
 \text{ over $c$-conjugate tuples} \ \bbf \in \cD^k.
\end{align}
As a result, the solution curve $\ga_{\bbf}$ becomes restricted to be monotone in $\Gamma_{\bbf}$ while having a minimal distance to the data. 

This penalty-based optimization categorizes our monotone curve-fitting task to the first branch in the curve literature discussed in Section~\ref{sec:literature_curve}. More specifically, the self-consistency curve $\ga_{\text{SC}}$ of \cite{hast:stue:89} can also be seen to minimize the reconstruction error $\E[ \| \bbX - \ga_{\text{SC}}(s) \|^2]$ in the sense that their approach is to find the projection index that has smaller reconstruction error $s_{\ga_{\text{SC}}}(\bbX):=\sup\{s: \| \bbX-\ga_{\text{SC}}(s) \| =\inf_{s'} \|\bbX-\ga_{\text{SC}}(s') \|\}$ where $\ga_{\text{SC}}(s)=\E[\,\bbX\,|\, s_{\ga_{\text{SC}}}(\bbX)=s\,]$. In our case, we specify the diagonal coordinate, which is not a variable of optimization, but still maintains the data-driven nature by summing all variables. Similarly, other methodologies in this branch formulate constrained optimization to define and find principal curves \citep{tibs:92,kegl:etal:00,haub:15}.

In conclusion, we propose the following monotone curve-fitting task.
\begin{algo}\label{algo} 
Given data $\widehat \mu \in \cP(\R^k)$, solve \eqref{problem3d} or \eqref{problem3d-1} to find a minimizer of convex functions $\widehat{\bbf} = (\widehat f_1,\dots,\widehat f_k)$. The parametrized monotone curve $\widehat\ga(s) := \big((\p \widehat g_i)^{-1} (s)\big)_{i \in [k]}$ is offered as a solution for the monotone curve-fitting task, where $\widehat g_i = \widehat f_i +q$. 
\end{algo}

\subsubsection{Extension of Task~\ref{algo} via orthogonal transformation}

What can we do if, for instance, the observed data in $\R^2$ does not exhibit rough monotonicity but instead aligns with an anti-monotone structure, such as along the anti-diagonal? In cases where the data appears to be roughly aligned along an axis that is not parallel to the diagonal, one could consider applying Task \ref{algo} after performing an appropriate {\em rotation of the data}. We denote by ${\cal O}(k) = \{ U \in \R^{k \times k} \, | \, U^\trans U = U U^\trans = I \}$ the set of all orthogonal matrices, 
where $U^\trans$ is the transpose of $U$ and $I$ is the identity matrix. Then for $\bbf = (f_1,\dots, f_k) \in \cD^k$ with a chosen domain $\cD \subset \cA(\R)$, the problems \eqref{problem3d} and \eqref{problem3d-1} can be generalized as

\begin{align}
&\text{minimize } \E_{\bbX \sim \mu} \big[H(U\bbX; \bbf) + \la  \| U\bbX - \ga_{\bbf} (s(U\bbX)) \|^2 \big]  \text{ over $U \in {\cal O}(k)$ and $\bbf$ in duality}, \label{problem3d-ortho} \\
&\text{minimize } \E_{\bbX \sim \mu} \big [H(U\bbX; \bbf) + \la \| U\bbX - \ga_{\bbf} (s(U\bbX)) \|^2  \big] \text{ over $U \in {\cal O}(k)$ and $c$-conjugate } \bbf, \label{problem3d-ortho-1}
\end{align}
respectively. Consequently, we propose the following modified task. 
\begin{algo}\label{algo2} 
Given data $\widehat \mu \in \cP(\R^k)$, solve \eqref{problem3d-ortho} or \eqref{problem3d-ortho-1} to find a minimizer of an orthogonal transformation $\wh U$ and convex functions ${\widehat  \bbf} = (\widehat f_1,\dots,\widehat f_k)$. The parametrized monotone curve $s \mapsto \wh U^\trans \widehat\ga(s)$, where $\widehat\ga(s)= \big((\p \widehat g_i)^{-1} (s)\big)_{i \in [k]}$ and $\widehat g_i = \widehat f_i +q$, is offered as a solution.
\end{algo}

To motivate this task, we illustrate Task~\ref{algo2} on a toy example in Figure~\ref{algo2}. A toy data set follows a bivariate normal variable $\bbX(s)=(X_1(s),X_2(s))^\trans$ where $\E[X_1(s)]=-s^2$, $\E[X_2(s)]=\log(s+1)$, $\Var(X_1(s))=\Var(X_2(s))=0.1$, and $\Cov(X_1(s),X_2(s))=0.09$ for $s \sim {\rm Unif}[0,3]$. Note that the ground truth curve of $\bbX$ is expressed by $(-s^2,\log(s+1))^\trans$. The original data points are positioned antidiagonally (left-top figure), which causes the transformation $U$ to align the transformed data points diagonally. On top of that, the contour of $H$ exposes the monotone set. Once the task finds the monotone curve, the estimated curve in the space of $U\bbX$ is rotated back to estimate the curve in the space of $\bbX$. Depending on a goal of an application, this final step may not be necessary if one wants to find a monotone curve on the space of $U\bbX$ or find $U$ to make variables monotonically related. A numeric algorithm to implement Task~\ref{algo2} appears in Section~\ref{sec:numeric} later, where we use the first principal components of $\bbX$ to determine the initial value of $U$ and optimize it with other variables for sufficient flexibility.
\begin{figure}[h!]
    \centering
    \includegraphics[width=0.80\linewidth]{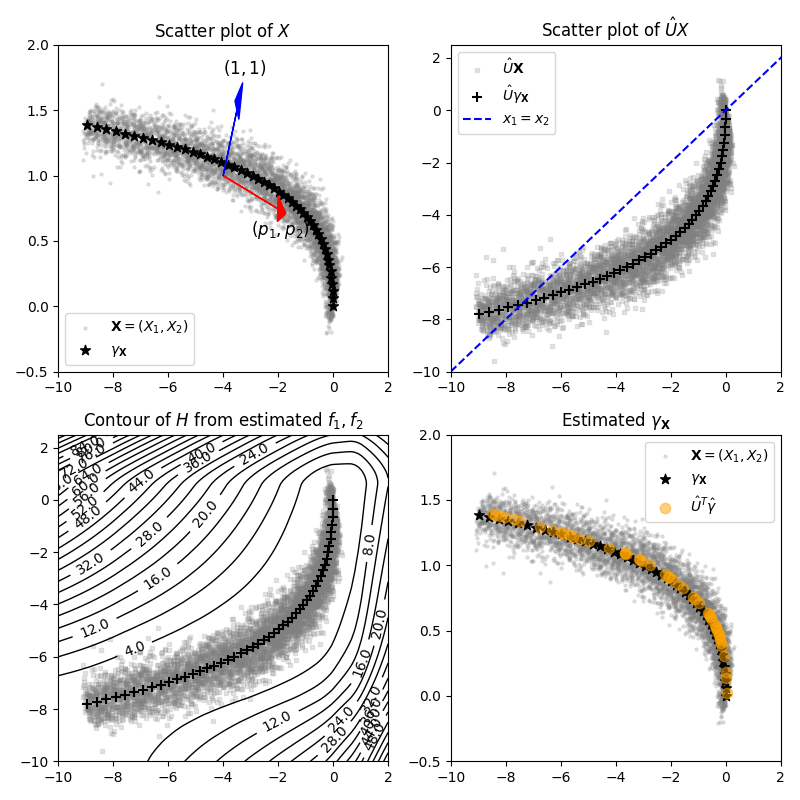}
    \caption{Task~\ref{algo2} for the toy example: $(p_1,p_2)$ is the first principal component, $(1,1)$ indicates the diagonal direction, and $\gamma_{\bbX}$ is the true curve associated to $\bbX$.}
    \label{fig:opt_overview}
\end{figure}

\section{Theoretical analysis}

In this section, we first show the existence of our curve estimate found from the statistical learning problem and then investigate statistical error bounds for the estimate. The proof and the explicit forms of technical constants appear in Supplementary \ref{supp:proof}. To highlight the theoretical contributions of our work, we first review key literature in the field. We use the asymptotic notation $a_n=O(b_n)$ (and $a_n=O_p(b_n)$) to imply that $a_n/b_n$ is bounded above by a constant (and with high probability), as $n$ gets larger for the sequences $a_n$, $b_n$.

In the first branch of principal curve literature (see Section~\ref{sec:literature_curve}), the work of \cite{kegl:etal:00}, assuming the curve has a fixed length on a bounded domain, showed that the rate of convergence is at $O(n^{-1/3})$, but the convergence is in terms of a loss function between data and a curve. That is, the loss function evaluating their empirical minimizer converges to the one of a true curve. More rigorous analysis of the statistical error rate stemmed from the second branch that defines the ridge of the target density as the principal curve. \cite{geno:etal:14} provided statistical foundations for \cite{ozer:etal:11}. They found a ridge estimator from kernel density estimation converges to the population ridge at $O_p((\log n / n)^{2/(d+8)})$ in the Hausdorff distance, where $d$ is the dimension of data. Moreover, they showed that if data is concentrated on a manifold with Gaussian noise having standard deviation $\sigma$ in the ambient space, and the ridge estimator aims to extract the manifold, then the rate has an additive error term $O(\sigma^2\log(1/\sigma))$, relying on the size of the noise, that does not decay by the sample size.

\label{sec:error_analysis}
\subsection{Existence}

To begin with, we show the existence of a solution to the problems \eqref{problem3d} and \eqref{problem3d-1} for any distribution of $\bX \sim \mu$ where it has the finite second moment.  
\begin{theorem}\label{existence}
Let $\cD = \cC_{\bbm}$ in \eqref{problem3d} and  \eqref{problem3d-1} where $\bbm = (m_0, m_1, m_2, m_3)$ with $m_i \ge 0$, and
\begin{align}\label{domain}
\cC_{\bbm} := \{f \in C^2(\R) : \ &|f(0)| \le m_0,\, |f'(0)| \le m_1,\, 0 \le f''(x) \le m_2,\\ &|  f''(x) - f''(x')| \le m_3 |x - x'| \, \text{ for all }\, x, x' \in \R\}.  \nn
\end{align}
If $\E_{\bbX \sim \mu} \| \bbX\|^2 < \infty$ and $m_2 \ge k-1$, then solutions to \eqref{problem3d} and \eqref{problem3d-1} exist, respectively. 
\end{theorem}
We remark that existing studies of a principal curve particularly for the first branch showed the existence of their principal curves under relatively stronger conditions. The principal curve with self-consistency condition in \cite{hast:stue:89} guarantees the existence for some specific distributions such as spherically symmetric and ellipsoidal distributions. In \cite{kegl:etal:00}, they showed the existence of a principal curve with a finite second moment, but they further require the curve to have a fixed length on a bounded domain. In contrast, our curve estimation enjoys the existence in more general distributions.

\subsection{Statistical error analysis}

In this section, we present our primary theoretical results on the statistical error bound of our estimate. 
We first outline the population model that generates the data. Consider a dimension \( k \geq 2 \), a maximally monotone curve \( \Gamma \subset \mathbb{R}^k \), and a probability distribution \( \rho \in \mathcal{P}(\mathbb{R}^k) \) such that \( \rho(\Gamma) = 1 \). Let \( \eps \in \mathcal{P}(\mathbb{R}^k) \) denote the noise distribution with zero mean. Define \( \bbU \) and \( \bbR \) as independent random variables in \( \mathbb{R}^k \), where \( \bbU \sim \rho \) and \( \bbR \sim \eps \), thus $\bbU \in \Ga$ and  \( \E [\bbR] = \textbf{0}  \). The population model is then described by \( \bbX = \bbU + \bbR \) with $\bX \sim \mu$, representing the population distribution. Let \(\gamma(s) = \big((\partial g_i)^{-1}(s)\big)_{i \in [k]}\) parametrize \(\Gamma\), where \(\bbf = (f_1, \dots, f_k)\) is $c$-conjugate and exposes $\Ga$, defined as in \eqref{potential3} in Supplementary~\ref{supp:extra}.

Denote \(\bbX = (X_1, \dots, X_k)\) and define \(\widehat \bbX := (\bbX_m)_{m=1}^n = (X_{1,m}, \dots, X_{k,m})\), \(m = 1, \dots, n\), as independent and identically distributed (i.i.d.) observations sampled from \(\mu\). The empirical data distribution is denoted by \(\widehat \mu =  \sum_{m=1}^n \delta_{ \bbX_m}/n\), where $\delta_x$ denotes the Dirac mass at $x$. Let  \(\widehat{\bbf} \in \cD^k\) be a $c$-conjugate solution to \eqref{problem3d-1} with $\mu$ replaced by \( \widehat \mu \), and let \(\widehat \gamma(s) = \big((\partial \widehat g_i)^{-1}(s)\big)_{i \in [k]}\) denote the corresponding parametrized monotone curve. Note that although $\widehat \mu$ and $ \widehat \gamma$  depend on the number of samples 
$n$, we omit it from the notations for simplicity. 

We first establish an upper bound for the {\em expected empirical MSE}, defined as \(E_{n}^{\textup{\textsf{emp}}} = \E\left[ n^{-1}\sum_{m=1}^n \| \ga (s(\bbX_m)) - \widehat{\ga} (s(\bbX_m)) \|^{2} \right]\), as stated in the following theorem. This result indicates that our curve estimate becomes more accurate as the noise in the data decreases.

\begin{theorem}\label{main}
Let $\bbm = (m_0,m_1,m_2,m_3) \in \mathbb R_{\ge 0}^4$ and $\cC_{\bbm}$ be the class of admissible functions as in \eqref{domain}.
Suppose $\Ga \subset \mathbb{R}^k $ is maximally and strictly monotone, and its associated $\bbf = (f_i)_{i \in [k]}$, defined in \eqref{potential3},  belongs to  $(\cC_{\bbm})^k$. Assume $\E \| \bbX\|^2 < \infty$. Given empirical data $\widehat \mu$, let $\widehat{\bbf}$ minimize \eqref{problem3d-1} with  $\cD = \cC_{\bbm}$. Then for any data dimension $k \ge 2$ and the number of samples $n \in \N$, we have $E_{n}^{\textup{\textsf{emp}}} \le C \E[ \| \bbR \|^2]$ where $C>0$ depends only on $m_2, \la$, and $k$. Moreover, if the components of the noise are mutually uncorrelated, i.e., $\E[R_i R_j] = 0$ for every $i \ne j$, then $C$ depends only on $m_2$ and $\la$, and not on $k$.
\end{theorem}

We note that this dependence of the empirical error bound on noise size aligns with the argument in \cite{hast:stue:89}, which discusses how noise size amplifies the bias of the self-consistent curve, particularly when the curve is an arc of a circle.

We then establish an upper bound for the {\em generalized MSE}, the estimation error for unseen data \(\bbY \overset{\text{d}}{=} \bbX\), where \(\bbY\) is independent of \(\bbX\), defined as \(E_{n}^{\textup{\textsf{gen}}} = \mathbb{E} \left[ \| \gamma(s(\bbY)) - \widehat \gamma(s(\bbY)) \|^{2} \right]\).
\begin{theorem}
\label{cor:gen_error}
Suppose the assumptions of Theorem~\ref{main} hold. If $\mu$ is compactly supported, then   $| E_{n}^{\textup{\textsf{gen}}}  - E_{n}^{\textup{\textsf{emp}}}|  \le C'k n^{-1/3}$
where  $C'>0$ depends only on $m_1, \kappa$ but not on $k,n$, where $\kappa$ is the smallest number such that $s(\bbX) \in [-\kappa, \kappa]$ a.s.. The generalized MSE therefore satisfies
\begin{equation*}
     E_{n}^{\textup{\textsf{gen}}}\le  C \E[ \| \bbR \|^2] + C' k n ^{-1/3}. 
\end{equation*}   
\end{theorem}
Interestingly, our analysis based on the mean squared error conceptually shares similarities with the results of \cite{geno:etal:14}, particularly for the manifold case. As their rate of convergence, the upper bound of our generalization error is also decomposed by a sample complexity and an additive term proportional to the size of underlying random noises scaled by a technical constant. In our case, the convergence rate is at $O(n^{-1/3})$ in the mean squared error. Also, we remark that our analysis directly tackles the error analysis between true and estimated curves in contrast to \cite{kegl:etal:00}, which advances the curve-level statistical analysis in the first branch of a principal curve literature.

\section{Numerical algorithm}
\label{sec:numeric}

In previous sections, we justify the loss function $\E_{\bbX \sim \mu} \big[H(U\bbX; \bbf) + \la  \| U\bbX - \ga_{\bbf} (s(U\bbX)) \|^2 \big]$ and its empirical version where $\bbf$ is in duality to eventually estimate the population monotone curve $\gamma$. This section introduces a numerical procedure to obtain the solution of Tasks \ref{algo} and \ref{algo2} to find $\widehat\ga(s) := \big((\p \widehat g_i)^{-1} (s)\big)_{i \in [k]}$ where $\widehat g_i = \widehat f_i +q$ and the diagonal coordinate $s$. One challenge stems from handling the inf-convolution $(\p \widehat{g}_i)^{-1} = \nabla ({\widehat f}_i^* \square q)$, since computing the convex conjugate of $\widehat{f}_i$ and performing the required operation together is generally nontrivial. 
To address this, we incorporate an auxiliary optimization procedure that directly computes the inverse of 
$\nabla (\widehat{f}_i + q)$. This inverse of gradient is denoted by $\widehat{G}^{\shortminus}_i(s):=( \p \widehat g_i)^{-1} (s)$, representing the target estimate as $\widehat\ga(s)=\big(\widehat{G}^{\shortminus}_i(s)\big)_{i\in [k]}$. Finally, an additional penalty is placed to enforce the inverse relationship between $\widehat{G}^{\shortminus}_i(s)$ and $\p \widehat g_i$, so the proposed algorithm bypasses to handle the inf-convolution. 

To approximate the target functions $f_i$ and $G_i^{\shortminus}$, we employ neural networks. Neural networks are composite functions constructed from multiple layers of affine transformations and activation functions. Their use in this context offers several advantages. First, neural networks possess a universal approximation property, enabling them to approximate any continuous function arbitrarily well \citep{devo:21}.  This aligns with the requirements of our framework, where $f_i$ and $\gamma$ are  continuous functions. Also, enforcing the convexity of $f_i$ is straightforward using structural modifications of neural networks.

Denote ${\bbG}_{\shortminus}(s):=(G_1^{\shortminus}(s),\dots,G_k^{\shortminus}(s))$ and $\wh G_i := \nabla \wh g_i = \nabla \wh f_i + \id$. With the consideration of the invertibility constraint for $G_i^{\shortminus}$, a proposed optimization to implement Task~\ref{algo2} is to solve
\begin{align}
\label{opt:main}
    \min_{{\bbf},\,{\bbG}^{\shortminus},\, U}\, \E_{\widehat{\mu}} [H(U\bbX; \bbf)] + \lambda \E_{\widehat{\mu}}[\| U\bbX - {{\bbG}_{\shortminus}}(s) \|^2] + \tau \sum_{i=1}^k \E_{\widehat{\mu}}[\,|\widehat{G}_i({G}_{i}^{\shortminus}(s))-s |^2],
\end{align}
where $s=\sum_{i=1}^k (U\bbX)_i$, $U \in {\cal O}(k)$, $f_i \in {\cal{NN}}_{\text C}$, and $G_i^{\shortminus} \in {\cal{NN}}_{\text F}$ for all $i=1,\dots,k$. Recall that the optimization requires two constraints $H(U\bbX; \bbf) \geq 0$ and $U^{\trans}U=UU^{\trans}=I$. In the case of $U=I$, the optimization reduces to solving Task~\ref{algo}. The third term in \eqref{opt:main} encodes the monotone structure inherent in $\nabla \wh g_i$, ensuring that the resulting curve, $\widehat{\bbG}_{\shortminus}$, remains monotone. For ${\cal{NN}}_{\text C}$ and ${\cal{NN}}_{\text F}$, we specify 4 hidden layers with 64 nodes and the last layer with 1 output node. The ELU activation function is imposed except for the last layer. The neural networks in ${\cal{NN}}_{\text C}$ additionally have the structure of the input convex neural network \citep{amos:etal:17} that concatenates the input to all hidden layers and imposes non-negativity constraints on the weight matrices.

Algorithm~\ref{alg:main} shows the overall optimization procedures in detail. The superscript $(t)$ denotes the $t$th iterate of it, $\bbX_i$ denotes the $i$th random sample, and $X_{j,i}$ denotes the $j$th component of $\bbX_i$. To handle the nonnegativity constraint for $H(U^{(t)}\bbX_i;\bbf^{(t)})$, this work adopts the Lagrangian dual formulation used in \cite{fior:etal:21} that is able to adapt neural networks. This Lagrangian formulation places an inner maximization problem $\lambda_L (0 - H(U^{(t)}\bbX; \bbf^{(t)}))$ for the variable $\lambda_L$. During the iterative computation, $\lambda_L$ gradually increases according to the violation of the constraint $\max\{-H(U^{(t)}\bbX_i;\bbf^{(t)}),0\}$. We also acknowledge putting a sufficiently large constant $\lambda_L$ for the constraint. In the algorithm, if $H(U^{(t)}\bbX_i;\bbf^{(t)})\geq 0$ for all $i$, $L^+$ is the unbiased estimator of $\E[H(U^{(t)}\bbX;\bbf^{(t)})]$. For stable updates of the orthogonal transformation matrix, we set $U^{(0)}$ as the inverse of the first PCA component $(p_1,\dots,p_k)^\trans$ of $\bbX$, so that the first principal component of the transformed data points $(U^{(0)}\bbX_i)_i$ in the first iteration align with the diagonal, as illustrated in Figure~\ref{fig:opt_overview}. The orthogonality constraint for $U^{(t)}$, i.e, $U^{\trans}U=UU^{\trans}=I$, also adopts the Lagrangian duality with the variable $\lambda_O$. Similar to $\lambda_L$, $\lambda_O$ increases such that the resulting transformation approximately satisfies the orthogonality. Algorithm~\ref{alg:main} reduces to the implementation of Task~\ref{algo} by fixing $U^{(t)}=I$ and ignoring the update for orthogonal transformation.

Algorithm~\ref{alg:main} adopts the early stopping rule. The $n$ size of random samples is randomly split by training $\widehat{\mu}_{\text{train}}=\{\bbX_{{i_t}}:i_t=1,\dots,n_{\text{train}}\}$ and validation data $\widehat{\mu}_{\text{val}}=\{\bbX_{{i}_v}:i_v=1,\dots,n_{\text{val}}\}$ with $n_{\text{train}} + n_{\text{val}}=n$. Then, in every iteration, models are updated based on $\widehat{\mu}_{\text{train}}$, but the algorithm terminates the training process if the validation error $\E_{\widehat{\mu}_{\text{val}}} [H(U^{(t)}\bbX; \bbf^{(t)})] + \lambda \E_{\widehat{\mu}_{\text{val}}}[\| U^{(t)}\bbX - {{\bbG}^{(t)}_{\shortminus}}(s) \|^2]$ no longer decreases. Theorem~\ref{cor:gen_error} justifies this training rule since the theory characterizes that the error gap between training and validation data should be negligible. In this work, the rule does not consider the inverse penalty to align with our theoretical argument. Note that numerous machine learning tasks frequently adopt this rule, as it helps to prevent a model from being overfitted to training data, so models trained on finite samples better represent the underlying population.
\begin{algorithm}[h]
\caption{Monotone Curve Estimation with Early Stopping}
\label{alg:main}
{\scriptsize
\begin{algorithmic}[1]
\State \textbf{Input:} Initialized neural networks $f_1^{(0)},\dots,f_k^{(0)},G_1^{\shortminus,(0)},\dots,G_k^{\shortminus,(0)}$; Initialize $U^{(0)}=\text{diag}(1/p_1,\dots,1/p_k)$; Set $\lambda, \lambda_S \geq 0$ and $\lambda_L^{(0)}=\lambda_O^{(0)}=0$; the learning rate $r$ for each component; $t=0$; $\text{val}_{\text{prev}}=\text{val}_{\text{current}}=\infty$ (or a sufficiently large number)
\State \textbf{Output:} $\widehat{\bbG}_{\shortminus}^*=(G_1^{\shortminus,(T)},\dots,G_k^{\shortminus,(T)})^\trans$, $U^{*} = U^{(T)}$, and $\bbf^*=(f_1^{(T)},\dots,f_k^{(T)})^\trans$

\While{$\text{val}_{\text{prev}} \geq \text{val}_{\text{current}}$}
    \State $\text{val}_{\text{prev}} \gets \text{val}_{\text{current}}$ 
    \State Set $s_{i_t} = \sum_{j=1}^k (U^{(t)}\bbX_{i_t})_{j}$ for all $i_t$
    \State $L^+ \gets n_{\text{train}}^{-1} \sum_{i_t=1}^{n_{\text{train}}} \max\{H(U^{(t)}\bbX_{i_t}; {\bbf}^{(t)}),0\}$ 
    \State $L^- \gets n_{\text{train}}^{-1} \sum_{i_t=1}^{n_{\text{train}}} \max\{-H(U^{(t)}\bbX_{i_t}; {\bbf}^{(t)}),0\}$ 
    \State $R \gets n_{\text{train}}^{-1} \sum_{i_t=1}^{n_{\text{train}}} \|U^{(t)}\bbX_{i_t} - \bbG^{(t)}_{\shortminus}(s_{i_t})\|^2$ 
    \State $M \gets n_{\text{train}}^{-1} \sum_{i_t=1}^{n_{\text{train}}} \sum_{j=1}^k |\widehat{G}_j^{(t)}({G}_{j}^{\shortminus,(t)}(s_{i_t}))-s_{i_t} |^2$ 

    \For{$j = 1$ to $k$}
        \State $f_j^{(t+1)} \gets f_j^{(t)} - r \frac{\partial}{\partial f_j} \left(L^+ + \tau M + \lambda^{(t)}_L L^- \right)$
        \State $G_j^{\shortminus,(t+1)} \gets G_j^{\shortminus,(t)} - r \frac{\partial}{\partial G_j^{\shortminus}} \left(\lambda R + \tau M \right)$
    \EndFor
    \State $\lambda^{(t+1)}_L \gets \lambda^{(t)}_L + r L^-$ 
    \State \textbf{ \% Update for Orthogonal Transformation \%}
    \State $P_O \gets \max\{\sum_{i,j}((U^{(t),T}U^{(t)} - I)^2 + (U^{(t)}U^{(t),T} - I)^2)_{i,j},0\}$
    \State $U^{(t+1)} \gets U^{(t)} - r \frac{\partial}{\partial U} \left(L^+ + \lambda R + \tau M + \lambda_O^{(t)} P_O \right)$
    \State $\lambda^{(t+1)}_O \gets \lambda^{(t)}_O + r P_O$
    \State \textbf{ \% Compute the validation metric \%}
    \State Set $s_{i_v} = \sum_{j=1}^k (U^{(t)}\bbX_{i_v})_{j}$ for all $i_v$
    \State $\text{val}_{\text{current}} \gets n_{\text{val}}^{-1} \sum_{i_v=1}^{n_{\text{val}}} \max\{H(U^{(t)}\bbX_{i_v}; {\bbf}^{(t)}),0\} + \lambda n_{\text{val}}^{-1} \sum_{i_v=1}^{n_{\text{val}}} \|U^{(t)}\bbX_{i_v} - \bbG^{(t)}_{\shortminus}(s_{i_v})\|^2$
    
    \State $t \gets t + 1$
\EndWhile
\State Set $T \gets t$
\end{algorithmic}
}
\end{algorithm}

\section{Simulation}
\label{sec:simulation}
We conduct a comparison study on experimental data sets to verify the performance of our monotone curve-fitting framework. Our study considers \cite{hast:stue:89} and \cite{ozer:etal:11} abbreviated by `HS' and `SCMS' (Subspace Constrained Mean Shift) respectively. These competing methods find general principal curves that are not necessarily monotone, and each approach has different statistical foundations for curve estimation. HS finds a curve that minimizes the mean-squared error on the notion of self-consistency, whereas SCMS finds the ridge of the probability density that is estimated by kernel density estimation. 

\subsection{Experiment data}
The $j$th experimental data in $\mathbb{R}^2$ is generated as follows,   
\begin{align}
\label{eqn:data_gen}
    \big(X^{(j)}_1(s),X^{(j)}_2(s)\big)^{\trans} \sim {\rm N}\big(\big(\mu^{(j)}_1(s),\mu^{(j)}_2(s)\big)^{\trans}, \big(\sigma^{(j)}_1(s),\sigma^{(j)}_{1,2}(s),\sigma^{(j)}_2(s)\big)^{\trans}\big),
\end{align}
where $s \sim S^{(j)}$ is assumed; $\big(\mu^{(j)}_1(s),\mu^{(j)}_2(s)\big)^{\trans}$ is the $j$th vector of mean functions w.r.t. $s$; $\sigma^{(j)}_1(s)$ and $\sigma^{(j)}_2(s)$ are the variance of each element while $\sigma^{(j)}_{1,2}(s)$ is the covariance between the elements. Note that the mean vector indicates the principal curve. Table~\ref{tab:exp_data} summarizes the parameters to generate experimental data sets. To collect data, we first generate $s \sim S^{(j)}$, and $\big(X^{(j)}_{1,i}(s),X^{(j)}_{2,i}(s)\big)^{\trans}$ is generated based on the evaluated function of $\mu$ and $\sigma$ for $i=1,\dots,5000$. Then the data is standardized such that each variable has zero mean and unit standard deviation before fitting the curves. This pre-processing step improves the performance of HS. 
\begin{table}[h]
\centering
\caption{Configuration of each experimental data in $\mathbb{R}^2$}
\label{tab:exp_data}
{\footnotesize
\begin{tabular}{ccccccc}
\hline
$j$ & $S^{(j)}$ & $\mu_1^{(j)}$ & $\mu_2^{(j)}$ & $\sigma_{1}^{(j)}$ & $\sigma_{2}^{(j)}$ & $\sigma_{1,2}^{(j)}$ \\
\hline
1 &{Unif}(-3,3) &  $\exp(s/10)+s$     &       $s^3/3 + s$       &           0.1     &        0.1        &       $0.1\times \min\{\cos(s\pi)\exp(|s|),1\}$                \\
2 &{Unif}(-3,3) &          $s$       &           $s$           &            0.1     &        0.1        &           $0.1\times \cos(s\pi)$                    \\
3 &{Unif}(0,3) &         $-s^2$      &       $\log(s+1)$     &              0.1     &        0.1        &   $0.09$                     \\
\hline
\end{tabular}
}
\end{table}

To also investigate the performance in $\mathbb{R}^3$, the above data-generating process  \eqref{eqn:data_gen} is further extended to sample the multivariate normal $\big(X^{(j)}_1(s),X^{(j)}_2(s),X^{(j)}_3(s)\big)^{\trans}$. This three-dimensional random vector inherits the parameters used in Table~\ref{tab:exp_data} for $\big(X^{(j)}_1(s),X^{(j)}_2(s)\big)^{\trans}$, and the parameters for $X^{(j)}_3(s)$ and associated dependencies appear in Table~\ref{tab:exp_data3d} where $\sigma_{3}^{(j)}$ is the standard deviation for $X_3^{(j)}$ and $\sigma_{l,m}^{(j)}$ is the covariance between $X_l^{(j)}$ and $X_m^{(j)}$.
\begin{table}[h]
\centering
\caption{Additional parameters for each experimental data in $\mathbb{R}^3$}
\label{tab:exp_data3d}
{\footnotesize
\begin{tabular}{cccccc}
\hline
$j$  & $\mu_3^{(j)}$ & $\sigma_{3}^{(j)}$  & $\sigma_{2,3}^{(j)}$ & $\sigma_{1,3}^{(j)}$  \\
\hline
1 &  $t$      &      0.1     &       0.05         &      $0.1\times \min\{\sin(t\pi)\exp(|t|),1\}$                   \\
2  & $t$      &      0.1     &        0.09         &       $0.1\times \sin(t\pi)$                                       \\
3  & $t$      &      0.1     &        0.09     &            0.09                 \\
\hline
\end{tabular}
}
\end{table}

\subsection{Comparison}
To begin with, the implementation of each method is briefly explained. The curves of the competing methods, HS and SCMS, are found through their open source libraries with the default configuration of their algorithms, where SCMS uses the Silverman's rule of thumb for the bandwidth parameter. For ours, the original data $\widehat{\mu}$ with 5000 instances is randomly split by 4500 and 500 to distinguish training $\widehat{\mu}_{\text{train}}$ and validation data $\widehat{\mu}_{\text{val}}$, then Algorithm~\ref{alg:main} is implemented. Since HS and SCMS find estimated curves using all data instances $\widehat{\mu}$, ours also returns the estimated curve by evaluating $U^{*,\trans}\widehat{\bbG}_{\shortminus}^{*}(s)$ on the entire data $\widehat{\mu}$. Multiplying $U^{*,\trans}$ produces the estimated curve in the original space. The choice of $\lambda$ and $\tau$ is discussed later. To see details of implementation, refer to Supplementary~\ref{supp:simul}.

\begin{figure}[h!]
    \centering
    \includegraphics[width=1.0\linewidth]{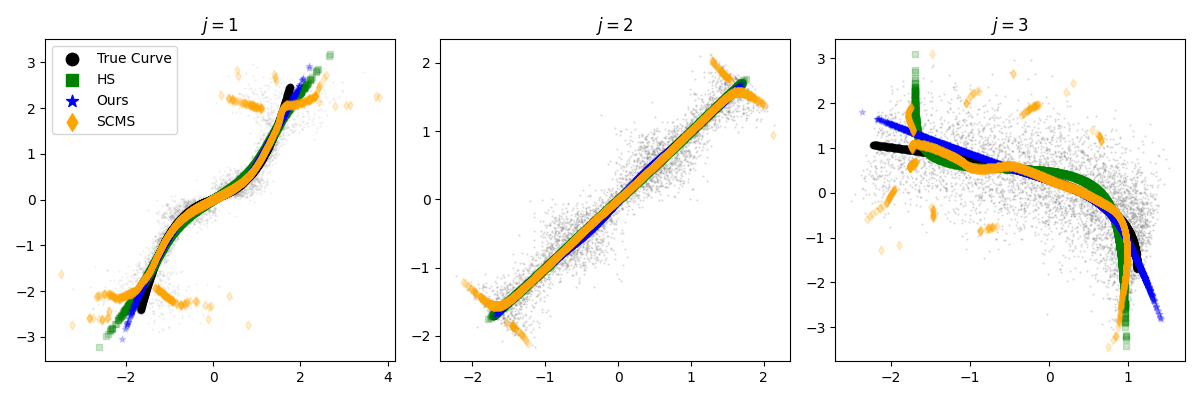}
    \includegraphics[width=1.0\linewidth]{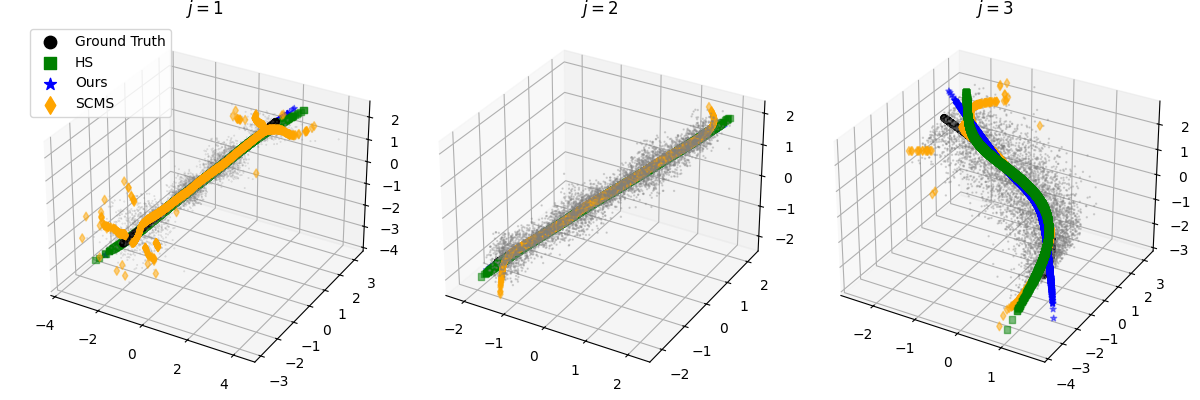}
    \caption{Comparison of three methods on 2 and 3 dimensions: the ground truth curves (black points), Ours (blue stars), SCMS (orange rhombuses), and HS (green rectangles).}
    \label{fig:toy_2dim}
\end{figure}

Visual inspection and quantitative evaluation support the superiority of our approach for finding the ground truth monotone curves. Figure~\ref{fig:toy_2dim} provides a visual comparison of the methods in both $\mathbb{R}^2$ and $\mathbb{R}^3$. HS and ours produce smooth curves; however, SCMS exhibits significant oscillations and erroneously traces data points as curves, particularly around the tails of the distributions. While HS appears to follow the monotone curves reasonably well in cases $j=1,2$, but for $j=3$, it abruptly bends upward near the top left region, significantly deviating from the true curve. Similar phenomena are also observed in the case of $\mathbb{R}^3$ as well. To evaluate the performance, we calculate averages and standard deviations of the Hausdorff and 2-Wasserstein distance between the estimated and the true curves, abbreviated by Haus. and Wass. respectively, based on 10 independent replicates for each method. All values are rounded up at the 3 decimal point and multiplied by 100 for clarity. Table~\ref{tab:toy_2dim} supports the visual inspections. Ours achieves better evaluation scores compared to the competing methods in general. Note that the Hausdorff distance is particularly useful for assessing the robustness of a method in noisy data.  The incorrect curve points of SCMS, which look isolated from the true curve, lead to substantially large Hausdorff distances.

\begin{table}[h!]
\centering
\caption{The number in the parenthesis stands for the standard deviation. Smaller values for each type of score are marked boldly.}
\label{tab:toy_2dim}
{\scriptsize
\begin{tabular}{ccccccccc}
\hline
&  & \multicolumn{2}{c}{Ours}      & \multicolumn{2}{c}{HS}        & \multicolumn{2}{c}{SCMS}  \\ \hline
$k$ & $j$    & Haus.($\downarrow$)         & Wass.($\downarrow$)              & Haus.($\downarrow$)         & Wass.($\downarrow$)              & Haus.($\downarrow$)           & Wass.($\downarrow$)     \\  \hline
\multirow{3}{*}{$\mathbb{R}^2$} & $1$ & {\bf 83.141 (10.270)} & {\bf 0.775 (0.137)} & 129.247 (17.985) & 1.420 (0.157)  & 205.784 (19.676) & 3.710 (0.305) \\
                                & $2$ & {\bf 7.499 (1.686)} & {\bf 0.196 (0.039)} & 15.856 (5.475) & 0.666 (0.129)  & 90.283 (14.467) & 1.013 (0.133) \\
                                & $3$ & {\bf 137.598 (27.392)} & {\bf 2.973 (0.397)} & 176.840 (25.085) & 8.655 (1.419)  & 224.469 (22.693) & 8.582 (0.918) \\ 
\hline
\multirow{3}{*}{$\mathbb{R}^3$} & $1$ & {\bf 74.966 (8.314)} & {\bf 0.689 (0.141)} & 115.325 (17.367) & 1.708 (0.165) & 215.840 (28.041) & 5.181 (0.479) \\
                                & $2$ & {\bf 30.062 (3.512)} & {\bf 0.150 (0.029)} & 47.032 (5.645) & 0.461 (0.059) & 91.037 (19.051) & 0.537 (0.051) \\
                                & $3$ & {\bf 147.316 (19.486)} & {\bf 4.588 (0.418)} & 197.863 (33.584) & 6.449 (0.493) & 248.761 (19.156) & 7.721 (0.920) \\
\hline
\end{tabular}
}
\end{table}

\subsubsection{Varying the size of noise} Table~\ref{tab:toy_noise_change} summarizes the evaluation scores for the different magnitudes of random noise whose variance and covariance components in \eqref{eqn:data_gen} are multiplied by $\sigma_f$, particularly for the case of $j=3$. 
 Note that the third experimental data involves more difficult curve estimation than the others intrinsically. In Table~\ref{tab:toy_2dim}, all methods show relatively higher errors for $j=3$. As shown in Table~\ref{tab:toy_noise_change}, the three models show improved performance as the noise scale gets smaller. In particular, we observe that ours is still consistently stronger in the Hausdorff distance but SCMS now becomes stronger in the 2-Wasserstein distance. Tables~\ref{tab:toy_2dim} and \ref{tab:toy_noise_change} highlight that our method tends to achieve more accurate estimation, especially in noisy and complex data exhibiting a roughly monotone shape.

\begin{table}[h!]
\centering
\caption{For the case of $j=3$, the variance and covariance components for $j=3$ are scaled by $\sigma_f$. The number in the parenthesis stands for the standard deviation, and the smaller values are marked boldly.}
\label{tab:toy_noise_change}
\vspace{0.5cm}
{\scriptsize
\begin{tabular}{ccccccccc}
\hline
 &  & \multicolumn{2}{c}{Ours}      & \multicolumn{2}{c}{HS}        & \multicolumn{2}{c}{SCMS}  \\ \hline
 & $\sigma_f$    & Haus.($\downarrow$)         & Wass.($\downarrow$)              & Haus.($\downarrow$)         & Wass.($\downarrow$)              & Haus.($\downarrow$)         & Wass.($\downarrow$)     \\  \hline
\multirow{2}{*}{$\mathbb{R}^2$}& $0.1$ & {\bf 38.675 (6.315)} & 0.919 (0.125) & 60.701 (18.866) & 0.298 (0.025) & 75.654 (12.892) & {\bf 0.273 (0.053) }\\ 
                               & $0.01$& {\bf 15.091 (1.947)} & 0.380 (0.054) & 17.164 (7.614) & 0.176 (0.016) & 18.963 (6.041) & {\bf 0.016 (0.002)} \\ 
\hline
\multirow{2}{*}{$\mathbb{R}^3$}& $0.1$ & {\bf 25.138 (4.638)} & 0.383 (0.096) & 52.253 (13.588) & 0.417 (0.043) & 75.009 (7.109) & {\bf 0.343 (0.035)} \\ 
                               & $0.01$& {\bf 12.350 (2.011)} & 0.235 (0.023) & 15.091 (5.152) & 0.289 (0.057) & 15.851 (4.286) & {\bf 0.017 (0.002)} \\ 
\hline
\end{tabular}
}
\end{table}

\subsection{Model selection}

This section discusses how to use the proposed framework in a principled way so that real-world users readily apply our method to their problems.

\subsubsection{Choice of $\lambda$ and $\tau$}
\label{simul:hyper}

Define $L_H = \E_{\widehat{\mu}_{\text{val}}} [H(U^*\bbX; \bbf^*)]$ and $L_R = \E_{\widehat{\mu}_{\text{val}}}[\| U^*\bbX - {\widehat{\bbG}^*_{\shortminus}}(s)\|^2]$.  Our proposed strategy is to select $\lambda$ and $\tau$ by minimizing $L_H + L_R$, the sum of the two main validation errors. Alternatively, other criteria can be considered, such as the weighted sum $w_1 L_H + w_2 L_R$ with $w_1+w_2=1$ or the Pareto frontier of $(L_H, L_R)$, depending on the specific domain knowledge or requirements. While metrics like Hausdorff and Wasserstein distances explicitly measure the estimation quality as the distance between the true and estimated curves, they are often impractical in real-world scenarios since the true curve is typically unknown.

Table \ref{tab:mc_ex3} summarizes the averages of the evaluation metrics for the 9 combinations of $\tau$ and $\lambda$ in the case of $j=3$. The table first highlights that our method achieves better Haus. scores compared to the competing methods in all the combinations of $\tau$ and $\lambda$. Secondly, using $\lambda$ and $\tau$ that achieves the minimum $L_H + L_R$ seems to avoid a relatively poor estimation. For example, as shown in the table, this tuning strategy avoids the choice $(\lambda, \tau) = (1,10)$ which has a relatively higher $L_H + L_R$. This choice exhibits a significantly worse Wass. score compared to other options in the table, as well as HS and SCMS in Table~\ref{tab:toy_2dim}. These findings motivate the adoption of the strategy proposed in this section, with the scores presented in Tables 3 and 4 being derived accordingly. We note that the results of $j=1$ and $j=2$ show a similar pattern, and are presented in Supplementary~\ref{supp:simul}.


\begin{table}[h!]
\centering
\caption{Evaluation metrics for different choices of $\lambda$ and $\tau$ when the data $j=3$ in $\mathbb{R}^2$: In each column of Wass. and Haus., and $L_H + L_R$, bold values indicate the minimum across combinations of $\lambda$ and $\tau$. Values marked with an asterisk ($\ast$) indicate the values of Wass. and Haus. that correspond to the minimum $L_H + L_R$.}
\label{tab:mc_ex3}
\vspace{0.5cm}
{\scriptsize 
\begin{tabular}{ccccccc}
\hline
 $\tau$ & $\lambda$       & Wass. & Haus. & $L_H$ &  $L_R$ &  $L_H+L_R$\\ \hline
\multirow{3}{*}{0.1}      & 1    &3.183 (0.420) &129.313 (22.703) & 30.114 (1.546)& 60.530 (1.617) & 90.645 (2.842)\\ 
                          & 10   &3.084 (0.600) &134.837 (22.196) & 32.189 (1.764)& 62.591 (1.225) & 94.779 (2.230)\\ 
                          & 100  & $\ast${\bf 2.973 (0.397)} & $\ast$137.598 (27.392) & 30.138(2.361) & 59.940(2.648)  & {\bf 90.078 (4.727)}\\ \hline
\multirow{3}{*}{1}        & 1    &3.960 (0.614) &100.764 (14.479) & 30.870 (2.889)& 62.139 (2.896) & 93.009 (5.675)\\ 
                          & 10   &3.280 (0.468) &127.209 (21.941) & 30.880 (3.949)& 60.554 (4.251) & 91.434 (8.174)\\ 
                          & 100  &3.310 (0.415) &137.856 (24.916) & 30.214 (2.876)& 61.174 (3.268) & 91.388 (5.876)\\ \hline
\multirow{3}{*}{10}       & 1    &8.476 (0.755)& 100.988 (10.723)& 35.569 (2.827)& 68.700 (2.625) & 104.269 (5.265)\\ 
                          & 10   &4.765 (0.512) & {\bf 98.660} (16.712) & 31.306 (4.148)& 61.704 (3.705) & 93.010 (7.779)\\ 
                          & 100  &3.275 (0.461) &132.662 (23.547) & 32.754 (3.470)& 61.122 (3.003) & 93.876 (6.249)\\ 
\hline
\end{tabular}
}
\end{table}

\subsubsection{Estimation with $U$}  
Another key consideration is whether to use the orthogonal transformation. Table~\ref{tab:orthogonal} compares evaluation scores when $U$ is a variable versus when it is fixed as $U=I$. As shown in the table, optimizing the transformation matrix significantly improves estimation accuracy, particularly when the underlying curve is non-increasing. The optimized transformation matrix repositions data points so that each variable exhibits an increasing relationship, allowing the algorithm to identify the increasing curve more effectively in the transformed space. However, if the original data already shows a clear increasing pattern, using the orthogonal transformation is not essential. For instance, in Table~\ref{tab:orthogonal}, only minor improvements are observed for $j=1$ and $j=2$ when optimizing $U$, and even with $U=I$, the results still outperform competing methods in Table~\ref{tab:toy_2dim} in terms of Haus. and Wass. distances.

\begin{table}[h!]
\centering
\caption{Comparison for using the orthogonal transformation in Algorithm~\ref{alg:main}}
\label{tab:orthogonal}
{\scriptsize
\begin{tabular}{ccccc}
\hline
      & \multicolumn{2}{c}{Variable $U$} & \multicolumn{2}{c}{Fixed $U=I$} \\ \hline
$j$      & Haus. ($\downarrow$)              & Wass. ($\downarrow$)             & Haus. ($\downarrow$)           & Wass. ($\downarrow$)           \\ \hline
1 & {\bf 83.141 (10.270)} & {\bf 0.775 (0.137)} & 86.543 (14.505) &  0.873 (0.152)  \\
2 & {\bf 7.499 (1.686)} & {\bf 0.196 (0.039)} &  8.264 (3.395) &  0.219 (0.074)  \\
3 & {\bf 137.598 (27.392)} & {\bf 2.973 (0.397)} &  181.608 (11.925) &  110.020 (4.215)  \\
\hline
\end{tabular}
}
\end{table}

\section{Real data application}
\label{sec:realdata}
In this section, we apply the proposed method to two noisy and complex real-world datasets where monotonicity is reasonably assumed or observed by general curve-fitting methods. To evaluate the robustness of our approach, we compare two estimated curves based on different types of variable transformations. Variable transformation, a common preprocessing step in machine learning, is used to stabilize training and meet statistical assumptions. Ideally, a robust curve estimation method should produce consistent results regardless of the chosen transformation technique, ensuring reliability in the decision-making process.

\begin{figure}[h!]
    \centering
    \subfloat[Prices (Pri.) are standardized (Stand.) after logarithmic transformation ($\log(\cdot)$) before estimating the curves.\label{fig:commodity_logsprice}]{%
        \includegraphics[width=0.75\linewidth]{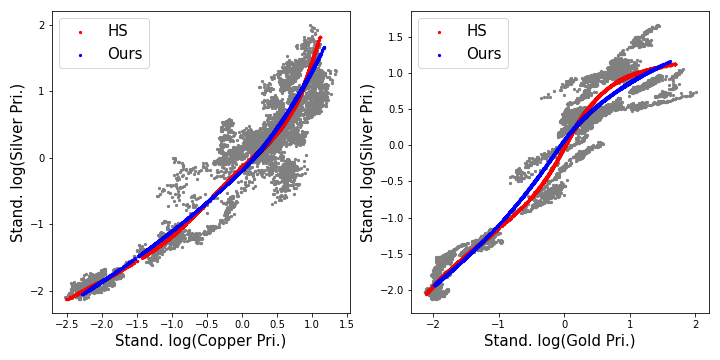}
    }
    
    \subfloat[Prices (Pri.) are standardized (Stand.) before estimating the curves.\label{fig:commodity_sprice}]{%
        \includegraphics[width=0.8\linewidth]{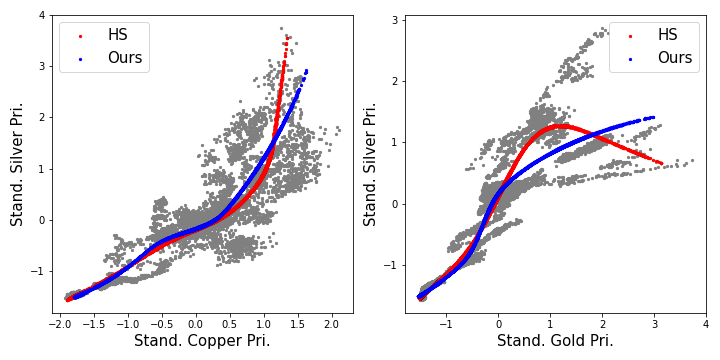}
    }
    
    \caption{Comparison of commodity prices: the prices of copper, silver and gold. Gray points are observed data points. Ours and HS curves are colored blue and red respectively.
    }
    \label{fig:commodity_2dim}
\end{figure}

\begin{figure}[h!]
    \centering
    \includegraphics[width=1.0\linewidth]{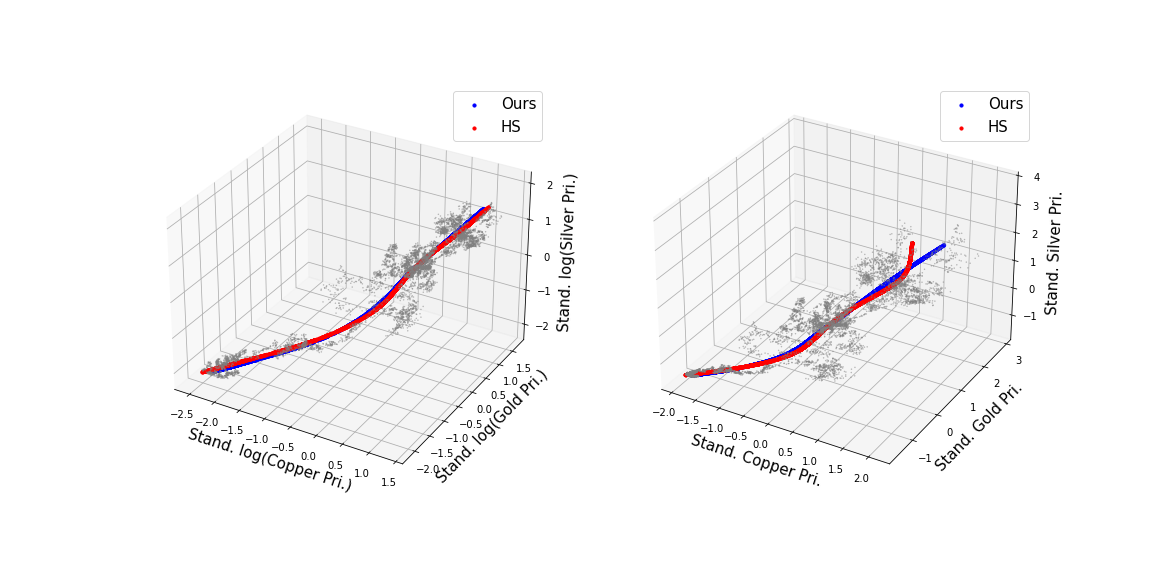}
    \vspace{-1.0cm}
    \caption{Comparison of commodity prices in $\mathbb{R}^3$: (Left) the prices are standardized after logarithmic transformation, and (right) the prices are just standardized before estimation.}
    \label{fig:commodity_3dim}
\end{figure}

\paragraph{Co-movement of commodity prices}

Analyzing commodity prices in terms of interdependence or reliance on economic indices is an important step toward gaining an in-depth understanding of global economics and financial markets. We apply the proposed method for modeling the future prices of copper, silver, and gold from August 30, 2000, to November 6, 2024. Each commodity price is retrieved via the Yahoo Finance API in R\footnote{quantmod::getSymbols(x, src = ``yahoo", from = ``2000-08-30", to = ``2024-11-06") \# Insert x=``HG=F" (``SI=F", ``GC=F", resp.).}. Rather than seeking economic insights, we attempt to address the advantage of enforcing monotonicity in curve estimation for robust inference by comparing the results to those of HS. To fit monotone curves, we use the same configuration as in the simulation section.


Figures~\ref{fig:commodity_2dim} and~\ref{fig:commodity_3dim} contrast the two estimated curves in $\mathbb{R}^2$ and $\mathbb{R}^3$ respectively. For the case of $\mathbb{R}^2$, we compare the price pairs: copper vs silver and gold vs silver. The case of $\mathbb{R}^3$ finds principal curves penetrating the middle of the three commodity prices. First, since the observed commodity prices tend to become more dispersed as the prices increase, the prices go through the logarithmic transformation and then standardization. Figure~\ref{fig:commodity_logsprice} displays the estimated HS curves monotonically increasing in both cases. This preceding procedure practically means that an HS curve can play the role of a profile analysis to justify using our monotone curve-fitting framework. Next, we find curves on data points only with standardization to see the impact of transformation on curve estimation. As shown in Figure~\ref{fig:commodity_sprice}, while our curve maintains the overall increasing shape, the shape of HS curve is inconsistent compared to Figure~\ref{fig:commodity_logsprice}. For the comparison between gold and silver prices, the HS curve does not represent the monotonic relationship anymore, which shows that the HS curve is sensitive to transforming variables. For the case of copper and silver in Figure~\ref{fig:commodity_sprice}, the downward curvature suddenly increases at 0.5 of \textsf{Stand. Copper Pri.}, pushing the curve sharply touches the point at its northernmost tip. This does not look representing the middle of data from our perspective. Figure~\ref{fig:commodity_3dim} compares the curves of all the commodities and observes the same phenomenon, addressing that ours enjoys the robustness in terms of variable transformation. 

\paragraph{Demand curve of avocado data}
The Avocado Prices and Sales Volume data\footnote{\url{https://www.kaggle.com/datasets/vakhariapujan/avocado-prices-and-sales-volume-2015-2023}} provides a comprehensive overview of avocado market trends in the United States over eight years between 2015 and 2023. The data includes important market information such as average prices, sales volume, and bag sizes for conventional and organic avocados in various regions of the United States. In this work, we are particularly interested in the joint relationship between prices (\textsf{AveragePrices}) and sales volume (\textsf{TotalVolume}) because these variables can be used to draw a demand curve for the avocado market. In economics, it is commonly believed that product prices and sales volume are inversely related because of the law of demand. We illustrate demand curves for organic avocados in San Francisco and Chicago. The curve of HS is also drawn for comparison. 
\begin{figure}[h!]
    \centering
    \includegraphics[width=0.8\linewidth]{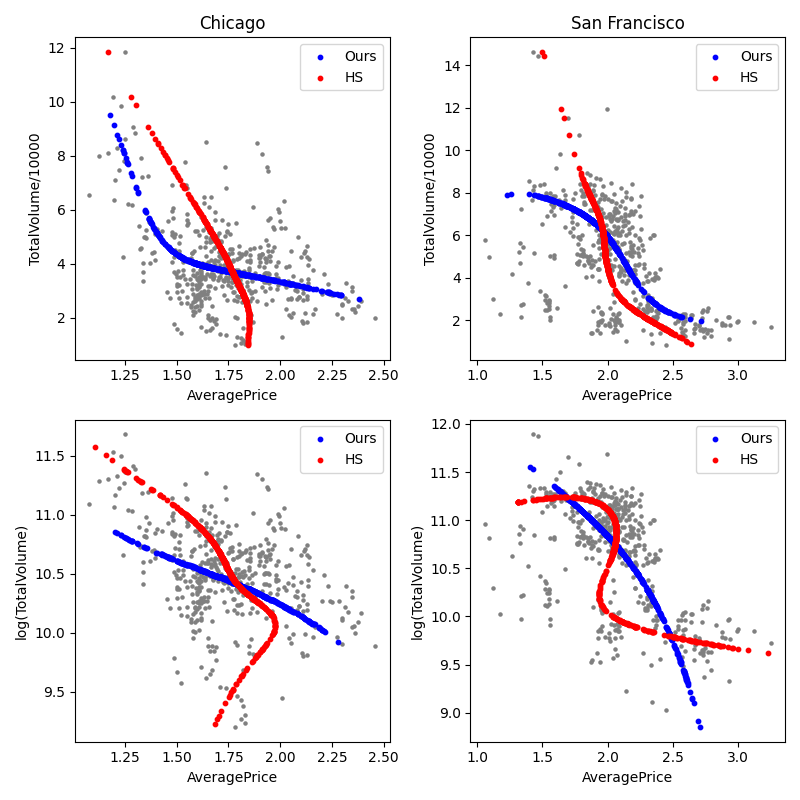}
    \caption{Demand curves: plots in the left are of Chicago and in the right of San Francisco.}
    \label{fig:avocado}
\end{figure}

Figure~\ref{fig:avocado} compares the demand curves of ours and HS. Because sales volume (\textsf{TotalVolume}) are on a large scale, we consider transforming it by \textsf{TotalVolume}/10000 and $\log (\textsf{TotalVolume})$. HS shows a monotonically decreasing pattern in the case of the linear transformation but nonmonotonic behaviors for the log transformation. In contrast, ours maintains a monotonically decreasing relationship regardless of the type of transformation. Although the curves of San Francisco have slightly different curvatures in higher prices, the overall monotonic shape remains. In general, when developing an operational strategy for a business, accurate and robust estimation of a demand curve is critical because it can be used to optimize pricing strategies and resource allocation in order to maximize potential revenue or profit.









\bibliographystyle{chicago}
\bibliography{ref.bib}

\clearpage
\pagenumbering{arabic}

\section{Additional discussion}
\label{supp:extra}

We further discuss the construction of a $c$-conjugate tuple disclosing a monotone set. This is an extension of Remark \ref{construction2} in the manuscript.
\begin{remark} 
\label{constructiond}
For simplicity (and without loss of generality), assume $\Ga \subset \R^k$ is maximally monotone. Let $\Ga_{ij} \subset \R^2$ represent the orthogonal projection of $\Ga$ onto the $x_ix_j$-plane. We further assume that $\Ga_{ij}$ is strictly monotone, and that the projections of $\Ga_{ij}$ to the $x_i$- and $x_j$-axes are both $\R$, for all $i < j$. As before, we identify $\Ga_{ij}$ with a function still denoted as $\Ga_{ij}$; $(x_i,  x_j) \in \Ga_{ij} \Leftrightarrow x_j = \Ga_{ij} (x_i)$. Let $\Ga_{ji} := \Ga_{ij}^{-1}$, and let $(0, y_{ij})$ denote the intersection point of $\Ga_{ij}$ with $x_j$-axis, $i < j$. Define
\begin{align}\label{potential3}
f_i(x_i) := \sum_{j=1}^{i-1} \int_{y_{ji}}^{x_i} \Ga_{ij}(u)du + \sum_{j= i+1}^k \int_0^{x_i} \Ga_{ij} (u)du.
\end{align}
Then, as explained in Remark \ref{construction2}, the following inequality holds:
\begin{equation*}
\sum_{i=1}^k f_i(x_i) = \sum_{i < j} \bigg[ \int_0^{x_i} \Ga_{ij} (u)du +  \int_{y_{ij}}^{x_j} \Ga_{ji}(u)du \bigg] \ge \sum_{i < j} x_ix_j \, \text{ for all } \,\bbx = (x_1,...,x_k) \in \R^k,
\end{equation*}
and moreover, equality holds if and only if $(x_i, x_j) \in \Ga_{ij}$ for all $i<j$, that is, $\bbx \in \Ga$. From this, it follows that the maximal monotonicity of $\Ga$ is equivalent to the $c$-conjugacy of $\bbf$.

Note that, defining $\Ga_{ii} := {\rm Id}$ the identity function on $\R$, $g_i = f_i + q$ can be expressed as 
\begin{equation*}
g_i(x_i) = \sum_{j=1}^{i-1} \int_{y_{ji}}^{x_i} \Ga_{ij}(u)du + \sum_{j= i}^k \int_0^{x_i} \Ga_{ij} (u)du.
\end{equation*}
\end{remark}

\section{Proof}
\label{supp:proof}

\subsection{Lemmas}

Here we present a few known facts with proofs for the sake of completeness.
\begin{lemma} \label{lemma1}
For $\bbf = (f_i)_{i=1}^k $ in duality, the set 
$\Gamma_{\bbf} = \{\bbx \in \R^k \, | \, H(\bbx; \bbf) = 0 \}$ is monotone. 
\end{lemma}
\noindent  \textit{Proof of Lemma \ref{lemma1}} Assume $\bbx = (x_i)_{i=1}^k, \, \bby = (y_i)_{i=1}^k \in \Gamma_{\bbf}$. For $c(\bbx) = \sum_{1 \le i < j \le k} x_i x_j$, it holds
\begin{align}\label{swab}
c(\bbx) + c(\bby) \ge c(x_1,...,x_{i-1},y_i, x_{i+1},...,x_k) + c(y_1,...,y_{i-1}, x_i, y_{i+1},...,y_k) \q \forall i \in [k],
\end{align}
since RHS $\le \big(\sum_{j \ne i} (f_j(x_j) + f_j(y_j))\big) + (f_i(y_i) + f_i(x_i)) = \sum_j (f_j(x_j) + f_j(y_j)) = c(\bbx) + c(\bby)$, where the equality holds since $\bbx, \bby \in \Ga_{\bbf}$. Let $d_i := x_i - y_i$, $i \in [k]$. Then \eqref{swab} is equivalent to 
\begin{align}\label{swab2}
d_i \cdot  \Big (\sum_{j \ne i} d_j\Big ) \ge 0, \q \forall i \in [k].
\end{align}
Without loss of generality, we may assume $d_1 > 0$. Then $\sum_{j \ge 2} d_j \ge 0$. If $d_2 < 0$, then \eqref{swab2} gives $d_1 + \sum_{j \ge 3} d_j \le 0$, but $d_1 + \sum_{j \ge 3} d_j = d_1 - d_2 + \sum_{j \ge 2} d_j \ge d_1 - d_2 > 0$, yielding a contradiction. We conclude that $d_i \ge 0$ for all $i \ne 1$, which proves the monotonicity of $\Gamma_{\bbf}$.

\begin{lemma} \label{coveringnum}
Let $I, J$ be compact intervals in $\mathbb R$. For any $\delta \in (0,1]$, the $\delta$-covering number $N_\delta$ for the set of $1$-Lipschitz functions from $I$ to $J$, with respect to the sup-norm, satisfies
    \begin{align*}
        \log N_\delta \le \frac{ |I|\log 3}{\de} + \log \Big(\frac{1}{\de}\Big) + \log ( 9|J| + 27).
    \end{align*}
\end{lemma}

\noindent \textit{Proof of Lemma \ref{coveringnum}}. We can let $I = [-a,a]$ and $J = [-b,b]$. Let $k,m \in \N$ satisfy $(k-1)\de < a \le k\de$ and $(m-1)\de < b \le m\de$. Let $I' := [-k\de, k\de]$, $J' := [-m\de, m\de]$. We discretize $I', J'$ by
    \begin{align*}
        V:= \{-k\de, -(k-1)\de,\dots, (k-1)\de, k\de\}, \ \  W:= \{-m\de, -(m-1)\de,\dots, (m-1)\de, m\de\}.
            \end{align*}
Let $S$ be the set of functions $\phi$ from $V$ to $W$ such that  for any  $k \in \mathbb Z$ with $k\delta , (k+1)\delta \in V$, 
\begin{align}\label{554}
\phi((k+1)\delta ) -   \phi(k \delta ) \in  \{-\delta,0,\delta\}.
\end{align}
Then we have
    \begin{align}\label{555}
  |S| \le |W| \cdot 3^{|V|-1} = (2m+1) \cdot 3^{2k}.
    \end{align}
For any $\phi \in S$, define its extension $\widetilde \phi : I' \to J'$ to be a linear interpolation of $\phi$. \eqref{554} implies $\widetilde \phi$ is $1$-Lipschitz. We claim that for any $1$-Lipschitz function $f: I' \to J'$, there exists $\phi \in S$ such that $||f-\widetilde \phi ||_\infty \le \delta$. To see this, we construct a function $\phi:V \rightarrow W$ as follows:
 \begin{align*}
       \phi(\delta n) := \text{the closest element in $W$ to } f(\delta n),
   \end{align*}
and if $f(\delta n) = \delta (m +\frac{1}{2})$ so there is a tie, then we set  $\phi(\delta n):= \delta m$. Since $f$ is  $1$-Lipschitz, $\phi$ satisfies \eqref{554} and thus $\phi \in S$. Finally, it is clear that $\| f-\widetilde \phi \|_\infty \le \de$. This implies $|N_\de| \le |S|$. 

By $k < \tfrac{a}{\de} +1$, $m < \tfrac{b}{\de} +1$ and $\de \le 1$, \eqref{555} gives
\begin{align*}
\log |S| &\le \log ( \tfrac{|J|}{\de} + 3) + ( \tfrac{ |I|}{\de} +2) \log3 \\
&\le  \log(\tfrac{1}{\de}) + \log (|J|+3) +   \tfrac{ |I|\log 3}{\de} +2 \log 3 \\
&\le \tfrac{ |I| \log 3 + \log(9 |J| + 27)}{\de}
\end{align*}
where we use $ \log({1}/{\de}) \le {1}/{\de} - 1$. Then $\log 2 \le {\log2}/{\delta}$ yields $\log (2|S|) \le  ({ |I| \log 3 + \log(18 |J| + 54)})/{\de}$.

\subsection{Theorem~\ref{existence}}
Let $L(\bbf) :=  \E_{\bbX \sim \mu} [H(\bbX; \bbf) + \la \| \bbX - \ga_{\bbf} (s(\bbX)) \|^2]$. For any $\bbf \in \cD^k$ in duality, we have $L(\bbf) < \infty$, since $H(\bbx; \bbf) \le c(1 + |\bbx|^2)$ for some $c > 0$, while $(\ga_{\bbf})_i = (\p g_i)^{-1}$ is $1$-Lipschitz. And if $\bbf = (k-1)(q,\dots,q)$, then $\bbf$ is $c$-conjugate and is in $\cD^k$ since we assume $m_2 \ge k-1$. Hence the problems \eqref{problem3d}, \eqref{problem3d-1} are well defined with nonempty domains, respectively.

Let $\bbf_n = (f_{1,n},\dots,f_{k,n})$ denote a minimizing sequence for the functional $L$, with each $\bbf_n \in \cD^k$ in duality. We show that one can extract a subsequence of $\bbf_n$ converging to a minimizer of $L$. By Arzelà–Ascoli theorem, there exists a subsequence of $(\bbf_n)_n$ (still denoted by $(\bbf_n)_n$) such that each $f_{i,n}$ and its derivatives $f'_{i,n}$, $f''_{i,n}$ converge uniformly on every compact subset of $\R$ to some function $f_i$ and its derivatives $f'_i$, $f''_i$, respectively, as $n \to \infty$. Thus $\bbf = (f_1,\dots,f_k)$ is in $\cD^k$, and the bound $0 \le H(\bbx; \bbf) + \la \| \bbx - \ga_{\bbf} (s(\bbx)) \|^2 \le C(1 + \| \bbx \|^2)$ allows us to apply the Lebesgue dominate convergence theorem to deduce $\lim_{n \to \infty} L(\bbf_n) = L(\bbf)$, showing $\bbf$ is a minimizer of $L$.\footnote{Here $C>0$ does not depend on $\bbf$; see \eqref{simplefact} where it is shown that $| \ga_i (0) | \le m_1$.} 

For the problem \eqref{problem3d-1}, we need to show  $S(\Ga_{\bbf}) = \R$, assuming $S(\Ga_{\bbf_n}) = \R$ for all $n \in \N$. \cite{bartz2021multi} showed that the maximality $S(\Ga_{\bbf}) = \R$ is equivalent to the condition 
\begin{align}\label{identity}
\prox_{f_1}+\cdots+\prox_{f_k}=  ( \p g_1)^{-1} +\cdots+   ( \p g_k)^{-1} =  \id \text{ on } \R,
\end{align}
 where $\displaystyle \prox_{f_i} (x) := \argmin_{p \in \R}\big( f_i(p)+| x-p |^{2}/2\big) = ( \p g_i)^{-1}(x)$ with $g_i = f_i +q$. 
 
 Fix $s \in \R$ and let $p_{i} := \prox_{f_{i}} (s)$ and $p_{i,n} := \prox_{f_{i,n}} (s)$. Then $p_{i}$ is the unique solution to the equation $g_{i}'(p) = s$. Since $1 \le g_{i,n}'' \le m_2 +1$ and $g_{i,n}'' \to g_{i}''$ locally uniformly, where $g_{i,n} = f_{i,n} + q$, we deduce $\lim_{n \to \infty} p_{i,n} =p_i$. This yields \eqref{identity}, thus concludes the proof.

\subsection{Theorem~\ref{main}}
We establish the upper bound of the expected empirical MSE. Recall $\widehat{\mu} = \sum_{m=1}^n \delta_{\bbX_m}/n$ denotes the empirical distribution given i.i.d. random samples $(\bbX_m)_{m=1}^n$,  $\widehat{\bbf} = (\widehat{f}_1,\dots, \widehat{f}_k)$ is an empirical  minimizer (i.e. $\widehat{\bbf}$ minimizes \eqref{problem3d-1} given $\widehat \mu$), and \(\widehat \gamma(s) = \big(( \widehat g'_i)^{-1}(s)\big)_{i \in [k]}\) is the resulting parametrized curve where $\widehat g_i = \widehat f_i +q$. Meanwhile, \(\gamma(s) = \big(( g'_i)^{-1}(s)\big)_{i \in [k]}\) parametrizes  \(\Gamma\), where \(\bbf = (f_1, \dots, f_k)\) is defined as in \eqref{potential3} and $g_i = f_i + q$.

\paragraph{Bound on the expected empirical MSE.} Since $\widehat{\bbf}$ is an empirical minimizer, we have
\begin{align} \label{obviousineq}
    &\frac{1}{n}\sum_{m=1}^n \Big[ \sum_{i=1}^k \widehat{g}_i(X_{i,m}) -\frac{1}{2} s(\bbX_m)^2  + \la \|  \bbX_m - \wh \ga \big(s(\bbX_m) \big)  \|^2 \Big] \nn \\
    &\q \le 
    \frac{1}{n}\sum_{m=1}^n \Big[ \sum_{i=1}^k {g}_i(X_{i,m})-\frac{1}{2} s(\bbX_m)^2  + \la \| \bbX_m -\ga \big(s(\bbX_m) \big) \|^2 \Big]
   \end{align}
(recall Remark \ref{fg}). We estimate an upper bound of the RHS. Since $f_i, \widehat f_i \in \cC_{\bbm}$, we have 
\begin{align}\label{2ndderivative}
1 \le g_i'' \le  m_2+1 \ \text{ and } \ 1 \le \widehat g_i'' \le m_2+1  \, \text{ for every } \, i \in [k].
\end{align}
Recalling the population model  \( \bbX = \bbU + \bbR \), we write $X_{i,m} = U_{i,m} + R_{i,m}$. By \eqref{2ndderivative},
\begin{align}
& \sum_{i=1}^k g_i(U_{i,m} + R_{i,m}) - \frac{1}{2} \Big(\sum_{i=1}^k (U_{i,m} + R_{i,m})\Big)^2 \nn \\
& \le  \sum_{i=1}^k \Big(g_i (U_{i,m}) + g_i' (U_{i,m})R_{i,m} + \frac{m_2+1}{2} R_{i,m}^2  \Big) - \frac{1}{2} \Big(\sum_{i=1}^k (U_{i,m} + R_{i,m})\Big)^2 \nn\\
& =  \sum_{i=1}^k g_i' (U_{i,m})R_{i,m} + \frac{m_2+1}{2}  \sum_{i=1}^k R_{i,m}^2 - \Big( \sum_{i=1}^k U_{i,m} \Big)\Big( \sum_{i=1}^k R_{i,m} \Big) - \frac12 \Big(\sum_{i=1}^k R_{i,m} \Big)^2 \label{goodupperbound'} \\
&\le  \sum_{i=1}^k g_i' (U_{i,m})R_{i,m} + \frac{m_2+1}{2}  \sum_{i=1}^k R_{i,m}^2 - \Big( \sum_{i=1}^k U_{i,m} \Big)\Big( \sum_{i=1}^k R_{i,m} \Big) \label{goodupperbound}
\end{align}
where the equality is due to $\bbU_m \in \Ga$ which yields $\sum_{i=1}^k g_i (U_{i,m}) - ( \sum_{i=1}^k U_{i,m})^2 /2 = 0$. Using the independence of  $\bbU_{m}$ and $\bbR_{m}$, $\E[ \| \bbR \|^2 ]= \sum_{i=1}^k \E[R_{i,m}^2] $, and $\E[\bbR]={\bf 0}$, we deduce 
\begin{align} \label{upperbound}
\E \bigg[\frac{1}{n}\sum_{m=1}^n \Big[ \sum_{i=1}^k {g}_i(X_{i,m}) -\frac{1}{2} \Big(\sum_{i=1}^k X_{i,m}\Big)^2 \Big]\bigg]
\le \frac{m_2+1}{2}\E[ \| \bbR \|^2 ].
\end{align}
If $\E[R_i R_j]=0$ for all $i \ne j$, then using \eqref{goodupperbound'}, we obtain the bound ${m_2}\E[ \| \bbR \|^2 ]/2$ in \eqref{upperbound}.

 Next, using $\bbU_{m} = \ga (s(\bbU_m))$ due to $\bbU_m \in \Ga$, and $s(\bbU_m) = s(\bbX_m) - s(\bbR_m)$, we compute
\begin{align}
\| \bbX_{m} - \ga (s(\bbX_m)) \|^2 &=  \| \bbU_{m} + \bbR_{m} -  \ga (s(\bbX_m)) \|^2 \nn \\
 &=  \| \ga (s(\bbX_m) - s(\bbR_m)) + \bbR_{m} -  \ga(s(\bbX_m)) \|^2 \nn \\
  &\le 2  \| \ga (s(\bbX_m) - s(\bbR_m))  -  \ga(s(\bbX_m)) \|^2 + 2 \|\bbR_{m}  \|^2 \nn \\
  &\le 2 | s(\bbR_m) |^2 +  2 \|\bbR_{m}  \|^2, \label{MSE1}
 \end{align}
where we use $\| \ga(s) - \ga(t) \| \le |s-t| $ for any $s,t \in \R$, which holds because, for any $s > t$,
\begin{align*}
\| \ga(s) - \ga(t) \|^2 = \sum_{i=1}^k ( \ga_i (s) - \ga_i (t) )^2 \le \Big( \sum_{i=1}^k (\ga_i (s) - \ga_i(t)) \Big)^2 = (s-t)^2
\end{align*}
where we use $\sum_{i=1}^k \ga_i = \id$ on $\R$, which is equivalent to the $c$-conjugacy of $\bbf$ (see \eqref{identity}).

Combining \eqref{goodupperbound}, \eqref{MSE1}, and the inequality $\E | s(\bbR_m)|^2 \le k \sum_i \E[R_{i,m}^2] = k \E[ \| \bbR \|^2 ]$, yield \begin{align}\label{Eupper1}
\E [ \text{RHS of $\eqref{obviousineq}$}] \le \big( \tfrac{m_2+1}{2} + 2\la (k+1) \big)\E[ \| \bbR \|^2 ].
\end{align}
If $\E[R_iR_j]=0$ for $ i \ne j$, then $\E | s(\bbR_m)|^2 =  \E[ \| \bbR \|^2 ]$, hence in this case, 
\begin{align}\label{Eupper2}
\E [ \text{RHS of $\eqref{obviousineq}$}] \le \big( \tfrac{m_2}{2} + 4\la  \big)\E[ \| \bbR \|^2 ].
\end{align}

We now estimate a lower bound of the LHS in \eqref{obviousineq}. For each $m=1,...,n$, we write
\begin{align} \label{three}
& \sum_{i=1}^k \widehat{g}_i(X_{i,m}) - \frac{1}{2} s(\bbX_m)^2 \nn 
    =  \Big[   \sum_{i=1}^k  \Big( \widehat{g}_i(X_{i,m})   - \widehat{g}_i ( \widehat{\ga}_i (s(\bbX_m))) \Big) \Big]\nonumber \\
 &\q\q\ + \Big[\sum_{i=1}^k  \widehat{g}_i ( \widehat{\ga}_i (s(\bbX_m))) - \frac{1}{2} \Big(\sum_{i=1}^k \widehat{\ga}_i (s(\bbX_m)) \Big  )^2  \Big]\nonumber  \\
&\q\q\ + \Big[ \frac{1}{2} \Big(\sum_{i=1}^k \widehat{\ga}_i (s(\bbX_m)) \Big)^2  - \frac{1}{2} s(\bbX_m)^2 \Big].
\end{align}
The second term in \eqref{three} is non-negative since $\sum_{i=1}^k \wh g_i (x_i) - (s(\bbx))^2/2\ge 0$ for any $\bbx \in \R^k$, while the third term vanishes due to the identity $\sum_{i=1}^k \wh \ga_i = \id$.  Meanwhile, using \eqref{2ndderivative}, we deduce that the first term in \eqref{three} is not smaller than
\begin{align*}
      \sum_{i=1}^k  &  \Big[ (\widehat{g}_i)'   ( \widehat{\ga}_i (s(\bbX_m)) )  ( X_{i,m} -  \widehat{\ga}_i (s(\bbX_m)) ) +
      \frac{1}{2}  ( X_{i,m} -  \widehat{\ga}_i (s(\bbX_m)))^2 \Big] \nonumber  \\
      &= \frac{1}{2} \sum_{i=1}^k  ( X_{i,m} -  \widehat{\ga}_i (s(\bbX_m)))^2 = \frac12 \| \bbX_{m} -  \widehat{\ga} (s(\bbX_m)) \|^2,
\end{align*}
where we used the fact $\wh \ga_i = ( \wh g_i ')^{-1}$  and  the identity $\sum_{i=1}^k \wh \ga_i = \id$. 

Using $\bbU_{m} = \ga (s(\bbU_m))$ and the inequality $\|a-b \|^2 \ge \|a \|^2/2 - \|b \|^2$, we compute
\begin{align*}
&\tfrac12 \| \bbX_{m} -  \widehat{\ga} (s(\bbX_m)) \|^2 = \tfrac12 \| \bbU_{m} + \bbR_{m} -  \widehat{\ga} (s(\bbX_m)) \|^2 \nn \\
 &= \tfrac12 \| \ga (s(\bbX_m) - s(\bbR_m)) + \bbR_{m} -  \widehat{\ga} (s(\bbX_m)) \|^2 \nn \\
  &= \tfrac12 \| [ \ga (s(\bbX_m)) -  \widehat{\ga} (s(\bbX_m))] - [  \ga (s(\bbX_m))   - \ga (s(\bbX_m) - s(\bbR_m)) - \bbR_{m} ] \|^2 \nn \\
  &\ge \tfrac14 \| \ga (s(\bbX_m)) -  \widehat{\ga} (s(\bbX_m)) ||^2 - \tfrac12 \|  \ga (s(\bbX_m))   - \ga (s(\bbX_m) - s(\bbR_m)) - \bbR_{m} \|^2 \nn \\
 &\ge  \tfrac14 \| \ga(s(\bbX_m)) -  \widehat{\ga}(s(\bbX_m)) \|^2 -   \|  \ga (s(\bbX_m))   - \ga(s(\bbX_m) - s(\bbR_m)) \|^2 -  \| \bbR_{m} \|^2 \nn \\
 &\ge  \tfrac14 \| \ga(s(\bbX_m)) -  \widehat{\ga} (s(\bbX_m)) \|^2 - |s(\bbR_m)|^2 - \| \bbR_{m} \|^2. 
\end{align*}
Combining the three lower bound estimates and  taking the expectation, we deduce
\begin{align} \label{lowerbound}
\E [ \text{LHS of $\eqref{obviousineq}$}] &\ge \frac{1}{n} \sum_{m=1}^n \Big(\frac12 + \la \Big) \E [ \| \bbX_m - \wh \ga \big(s(\bbX_m) \big) \|^2 ] \nn \\
& \ge  \frac{2\la + 1}{n} \sum_{m=1}^n \Big( \frac14 \E [\| \ga(s(\bbX_m)) -  \widehat{\ga} (s(\bbX_m)) \|^2] - (k+1) \E[ \| \bbR \|^2 ] \Big) \nn \\
&= \frac{2\la + 1}{4} E_{n}^{\textup{\textsf{emp}}} - (2\la +1)(k+1) ) \E[ \| \bbR \|^2 ]. 
\end{align}
In addition, if $\bbR$ has mutually uncorrelated components, due to $\E | s(\bbR_m)|^2 =  \E[ \| \bbR \|^2 ]$,
\begin{align} \label{lowerbound'}
\E [ \text{LHS of $\eqref{obviousineq}$}] \ge  \frac{2\la + 1}{4} E_{n}^{\textup{\textsf{emp}}} - 2(2\la +1) \E[ \| \bbR \|^2 ].
\end{align}
As a result, \eqref{obviousineq}, \eqref{Eupper1} and \eqref{lowerbound}  (or \eqref{obviousineq}, \eqref{Eupper2} and \eqref{lowerbound'}) yield an upper bound for $E_{n}^{\textup{\textsf{emp}}}$:
\begin{align}\label{empbound}
 E_{n}^{\textup{\textsf{emp}}} &\le \frac{(16\la +4)k + 16\la +2m_2 + 6}{2\la +1}\, \E[ \| \bbR \|^2], \q \text{and moreover,}  \nn \\
 E_{n}^{\textup{\textsf{emp}}} &\le \frac{32\la + 2m_2 + 8}{2\la +1}\, \E [ \| \bbR \|^2] \q \text{if } \  \E[R_i R_j] = 0 \, \text{ for every }  i \ne j.
\end{align}

\begin{remark} 
Assumption ii), i.e., $\bbf \in (\cC_{\bbm})^k$,  implies that the monotone functions $\Ga_{ij}$ in \eqref{potential3} for each $i \ne j$ is $C^1$ smooth with positive lower and upper bound on its derivative: $1/m_2 \le \Ga_{ij}'(x) \le m_2$ for all $i \ne j$ and $x \in \R$. To see this, observe that \eqref{potential3} implies
\begin{align}\label{potential'}
f_i'(x_i) = \sum_{j=1}^{i-1} \Ga_{ij}(x_i) + \sum_{j= i+1}^k \Ga_{ij}(x_i) \implies 
f_i''(x_i) = \sum_{j=1}^{i-1} \Ga_{ij}'(x_i) + \sum_{j= i+1}^k \Ga_{ij}'(x_i).
\end{align}
$\Ga_{ij}$ being $C^1$ for all $i \ne j$ is implied by their monotonicity and the assumption $f_i \in C^2(\R)$. Now \eqref{potential'} implies $\Ga_{ij}' \le m_2$ from $f_i'' \le m_2$, which in turn impies $\Ga_{ij}' \ge 1/m_2$ since $\Ga_{ji} = \Ga_{ij}^{-1}$.
\end{remark}

\subsection{Theorem~\ref{cor:gen_error}}

This section provides proof of the estimation gap and the generalized MSE in order. We suppose the assumptions used in Theorem~\ref{main} hold.
\paragraph{Gap estimation between $E_{n}^{\textup{\textsf{gen}}} $ and $E_{n} ^{\textup{\textsf{emp}}}$.} Let $(\bbY_m)_{m=1}^n$ be i.i.d. random vectors with \(\bbY_m \overset{\text{d}}{=} \bbX \sim \mu\) and $(\bbY_m)_{m=1}^n$ are independent of all other random variables. Then we write
\begin{align}\label{difference}
 &E_{n}^{\textup{\textsf{gen}}}  -  E_{n}^{\textup{\textsf{emp}}}  =\E\bigg[ \frac{1}{n}\sum_{m=1}^n \Big[ \| \ga (s(\bbY_m)) -  \widehat{\ga} (s(\bbY_m)) \|^2  - \| \ga (s(\bbX_m)) -  \widehat{\ga} (s(\bbX_m)) \|^2 \Big] \bigg] \nn \\
 &= \E\bigg[ \frac{1}{n}\sum_{m=1}^n \sum_{i=1}^k \Big[ | \ga_i (s(\bbY_m)) -  \widehat{\ga_i} (s(\bbY_m)) |^2  - | \ga_i (s(\bbX_m)) -  \widehat{\ga_i} (s(\bbX_m)) |^2 \Big] \bigg].
\end{align}
Let $\De_i := \ga_i - \wh\ga_i$. Note that, for any $\vp \in C(\R)$ and $m \ge 0$, the following holds:
\begin{align}\label{simplefact}
\text{If } |\vp(0)| \le m \text{ and } \vp(y) - \vp(x) \ge y-x \text{ for any } y > x, \text{ then } |\vp^{-1}(0)| \le m,
\end{align}
since $|\vp^{-1}(0)| \le | \vp(\vp^{-1}(0)) - \vp(0)| \le m$. Now recall $\ga_i = (g'_i)^{-1} = (f'_i + \id)^{-1}$ for some $f_i \in \cC_{\bbm}$. Applying \eqref{simplefact} with $\vp = g_i',\, m = m_1$  implies $| \ga_i (0) | \le m_1$, and similarly $| \wh\ga_i (0) | \le m_1$. Hence, $|\De_i (0)| \le 2m_1$. Also recall that $\ga_i$ and $  \wh \ga_i$ are nondecreasing 1-Lipschitz functions, implying that $\De_i$ is 1-Lipschitz on $\R$.

Since $\mu$ is compactly supported, there exists $\kappa \ge 0$ such that $s(\bbX) \in I := [-\kappa, \kappa]$ almost surely. Define $J = [-2m_1 - \kappa, 2m_1 + \kappa]$ so that any 1-Lipschitz function $\vp$ on $\R$ with $| \vp(0) | \le 2m_1$ satisfies $\vp(x) \in J$ for all $x \in I$. Define the set of functions $\cL = \{ \vp : I \to J \, | \, \vp \text{ is 1-Lipschitz} \}$, equipped with the sup norm $\|  \vp \|_\infty := \sup_{x \in I} |\vp(x)|$. Note that $\De_i \in \cL$ (if restricted on $I$) for all $i \in [k]$. Since $\cL$ is compact by Arzelà–Ascoli theorem, for any $\de > 0$, there exists $ N_\de \in \N$ and functions $w_1,...,w_{N_\de} \in \cL$ such that for any $\vp \in \cL$, $ \| \vp - w_j \|_\infty \le \de$ for some $j \in [N_\de]$. In fact, we have
     \begin{align}\label{entropy}
        \log 2N_\delta \le \tfrac{ |I| \log 3 + \log (18 |J| + 54)}{\de} \q \text{for any } \ \delta \in (0,1]. 
    \end{align}
See \cite{covering1,covering2} for the references and Lemma \ref{coveringnum} in Appendix for the proof.

Now observe that, for each $i \in [k]$, there exists a random index $r_i \in [N_\de]$ such that 
\begin{align*} 
    \| \De_i - w_{r_i} \|_\infty \le \de.
\end{align*}
The randomness of $r_i$ stems from the randomness of $\wh\ga$, which depends on random samples $(\bbX_m)_{m=1}^n$. Define  $h_\ell(\textbf{x},\textbf{y}):=| w_\ell (s(\bby) ) |^2 - |w_\ell (s(\bbx) ) |^2$, $\ell \in [N_\de]$. We can bound \eqref{difference} as
 \begin{align}  
     & |  E_{n}^{\textup{\textsf{gen}}}  - E_{n}^{\textup{\textsf{emp}}}| \nn \\
     & \le \bigg\vert \mathbb E  \bigg[  \frac{1}{n}\sum_{m=1}^n\sum_{i=1}^k h_{r_i}(\bbX_m,\bbY_m) +  \frac{1}{n}\sum_{m=1}^n\sum_{i=1}^k \Big( \De_i (s(\bbY_m))^2 - \De_i (s(\bbX_m))^2 - h_{r_i}(\bbX_m,\bbY_m) \Big)  \bigg]  \bigg\vert \nn \\
&\le  \bigg\vert \mathbb E  \bigg[  \frac{1}{n}\sum_{m=1}^n\sum_{i=1}^k h_{r_i}(\bbX_m,\bbY_m) \bigg] \bigg\vert  +4kM\delta, \label{genempdiff}
 \end{align}
where $M := 2m_1 + \kappa$, which yields $\| \De_i + h_{r_i} \|_\infty \le 2M$, hence $\| \De_i^2 - h_{r_i}^2 \|_\infty \le 2M\de$.

For any \emph{fixed} index $\ell \in [N_\de]$, since  $\{h_\ell (\textbf{X}_m,\textbf{Y}_m)  \}_{m=1,...,n}$ are  i.i.d. uniformly bounded (by $M^2$) and centered  random variables, Hoeffding's concentration inequality implies 
\begin{align*}
    \mathbb{P}\Big( \Big\vert\sum_{m=1}^n h_\ell (\textbf{X}_m,\textbf{Y}_m) 
 \Big\vert  \ge t\Big) \le 2 e^{-t^2/2nM^4}.
\end{align*}
This, combined with a union bound $\mathbb{P}(\bigcup_{j=1}^\infty A_j) \le \sum_{j=1}^\infty \mathbb{P}(A_j)$ for any events $(A_j)_j$, implies
\begin{align*}
     \mathbb{P}\Big( \Big\vert\sum_{m=1}^n h_{r_i} (\textbf{X}_m,\textbf{Y}_m)   \Big\vert  \ge t\Big)   \le \min \{1, 2N_\delta e^{-t^2/2M^4n} \} \q \text{for any }\, i \in [k].
\end{align*}
Hence, for any $u > 0$, we have
\begin{align*}
    \mathbb{E}\Big\vert\sum_{m=1}^n h_{r_i} (\textbf{X}_m,\textbf{Y}_m)   \Big\vert &=  \int_0^\infty \mathbb{P}\Big( \Big\vert\sum_{m=1}^n h_{r_i} (\textbf{X}_m,\textbf{Y}_m)   \Big\vert  \ge t\Big) \, dt\\
    & \le \int_0^u  dt + 2N_\delta \int_u^\infty e^{-t^2/2M^4 n} dt  \\
& \le u + 2N_\delta M^2 \sqrt{2\pi n}\, e^{-u^2/2M^4 n},
\end{align*} 
where we use a Gaussian tail bound $ \int_u^\infty e^{- {t^2}/{2\si^2}} dt/\sqrt{2\pi \si^2} \le e^{-u^2/2\si^2}$. 

Choosing $u$ to solve $2N_\delta e^{-u^2/2M^4 n} = 1 \Leftrightarrow u^2 = 2M^4 n \log (2N_\de)$, we deduce 
\begin{align*}
\mathbb{E}\Big\vert\frac1n\sum_{m=1}^n h_{r_i} (\textbf{X}_m,\textbf{Y}_m)   \Big\vert  \le M^2 \sqrt{2/n} \big( \sqrt{\pi}  + \sqrt{\log 2N_\de }\big).
\end{align*} 
This, along with \eqref{genempdiff}, \eqref{entropy} and taking  $\delta  = n^{-1/3} $ yields 
\begin{align}\label{CC}
| E_n^{\textup{\textsf{gen}}}  - E_n^{\textup{\textsf{emp}}}| \le kM^2(2\pi)^{1/2} n^{-1/2} + kM^2 (2C)^{1/2} n^{-1/3} + 4kM n^{-1/3}, 
\end{align}
where $C= |I| \log 3 + \log(18 |J| + 54)$. Using $|I| = 2\kappa$, $|J| = 2(2m_1 + \kappa)$, $\delta \le 1$, and $n^{-1/2} \le n^{-1/3}$, from \eqref{CC}, we can deduce the bound 
$|  E_{n}^{\textup{\textsf{gen}}}  - E_{n}^{\textup{\textsf{emp}}}|  \le C' k n^{-1/3}$ in Theorem~\ref{cor:gen_error} 
with
\begin{align}\label{C'}
C' = (2m_1 + \kappa) \big( \sqrt{2} (2m_1 + \kappa) \big( \sqrt{\pi} + \sqrt{2\kappa \log 3 + \log (36(2m_1 + \kappa) + 54)}\,\big) + 4 \big).
\end{align}
This completes the proof.

\section{Simulation}
\label{supp:simul}

\subsection{Implementation of competing methods}
The open source libraries of each method appear in the footnotes of HS\footnote{The R package is available at https://cran.r-project.org/web/packages/princurve/.},  SCMS\footnote{The Python library is available at https://github.com/zhangyk8/EuDirSCMS.}, and ours\footnote{\url{https://anonymous.4open.science/r/mono_curve/README.md}}.

\subsection{Additional tables}
This section includes Tables~\ref{tab:mc_ex1} and \ref{tab:mc_ex2} that summarize the scores for the choice of different $\lambda$ and $\tau$. With Table~\ref{tab:mc_ex3} in the manuscript, these additional tables support the discussion in Section~\ref{simul:hyper} about the hyperparameter tuning strategy.
\begin{table}[ht!]
\centering
\caption{Evaluation metrics for different choices of $\lambda$ and $\tau$ when the data $j=1$ in $\mathbb{R}^2$: Refer to Table~\ref{tab:mc_ex3} for details.}
\label{tab:mc_ex1}
\vspace{0.5cm}
{\scriptsize
\begin{tabular}{ccccccc}
\toprule
 $\tau$ & $\lambda$       & Wass. & Haus. & $L_H$ &  $L_R$ &  $L_H+L_R$\\ \midrule
\multirow{3}{*}{0.1}      & 1    & 1.005 (0.100)  & 88.087 (11.637) & 7.672 (1.667) & 25.679 (1.978) & 33.351 (3.560) \\
                          & 10   & 0.637 (0.108)	& {\bf 77.970 (12.575)} & 7.268 (1.526) & 24.511 (2.039) & 31.779 (3.449) \\
                          & 100  & {\bf 0.555 (0.096)} & 78.474 (12.427) & 8.075 (1.312) & 24.860 (1.096) & 32.935 (1.782) \\ \midrule
\multirow{3}{*}{1}        & 1    & 2.844 (0.431) &103.487 (9.538) & 8.010 (2.287) & 27.490 (1.894) & 35.500 (4.015) \\
                          & 10   & $\ast$0.775 (0.137) & $\ast$83.141 (10.270) & 7.008 (1.101) & 24.327 (2.308) & {\bf 31.335 (3.222)} \\
                          & 100  & 0.561 (0.077)	& 78.993 (13.812) & 7.063 (1.454) & 24.319 (1.612) & 31.382 (2.673) \\ \midrule
\multirow{3}{*}{10}       & 1    & 3.997 (0.155)  & 90.060 (12.504) & 8.732 (1.620) & 29.641 (2.523) & 38.373 (4.037) \\
                          & 10   & 2.658 (0.896)	& 97.116 (12.473) & 9.489 (1.986) & 27.004 (2.554) & 36.493 (4.370) \\
                          & 100  & 0.661 (0.098)	& 83.194 (13.019) & 7.849 (1.630) & 24.202 (2.029) & 32.051 (3.342) \\
\bottomrule
\end{tabular}
}
\end{table}

\begin{table}[ht!]
\centering
\caption{Evaluation metrics for different choices of $\lambda$ and $\tau$ when the data $j=2$ in $\mathbb{R}^2$: Refer to Table~\ref{tab:mc_ex3} for details.}
\vspace{0.5cm}
\label{tab:mc_ex2}
{\scriptsize
\begin{tabular}{ccccccc}
\toprule
 $\tau$ & $\lambda$       & Wass. & Haus. & $L_H$ &  $L_R$ &  $L_H+L_R$\\ \midrule
\multirow{3}{*}{0.1}      & 1    & 0.185 (0.039) & 7.527 (2.721)  &3.259 (0.473) & 20.785 (1.028) & 24.044 (1.359)\\ 
                          & 10   & 0.181 (0.032) & 7.765 (2.623)  &4.175 (0.744) & 21.221 (1.447) & 25.396 (1.621)\\ 
                          & 100  & 0.168 (0.037) & 6.943 (2.792)  &3.869 (0.820) & 21.046 (0.986) & 24.915 (1.553)\\ \midrule
\multirow{3}{*}{1}        & 1    & {$\ast$0.196 (0.039)} & {$\ast$7.499 (1.686)}  &3.156 (0.710) & 19.608 (1.797) & {\bf 22.764 (2.322)}\\ 
                          & 10   & {\bf 0.166 (0.020)} & {\bf 6.931 (2.896)}  &4.269 (0.650) & 21.838 (1.165) & 26.107 (1.520)\\ 
                          & 100  & 0.187 (0.036) & 7.629 (2.946)  &3.911 (0.748) & 21.187 (1.150) & 25.098 (1.317)\\  \midrule
\multirow{3}{*}{10}       & 1    & 0.379 (0.243) & 9.924 (3.540)  &3.917 (0.653) & 19.537 (1.690) & 23.454 (2.119)\\ 
                          & 10   & 0.211 (0.038) & 7.727 (2.130)  &4.728 (0.721) & 19.674 (1.184) & 24.402 (1.658)\\ 
                          & 100  & 0.198 (0.028) & 8.227 (2.353)  &5.556 (1.423) & 20.355 (1.283) & 25.912 (2.229)\\ 
\bottomrule
\end{tabular}
}
\end{table}




\bibliographystyleSupp{chicago}

\end{document}